\documentclass[oldversion]{aa}
\usepackage{graphicx}
\usepackage{txfonts}
\usepackage{natbib}
\bibpunct{(}{)}{;}{a}{}{,}
\begin{document}

\title{Reconstructing the intergalactic UV background with QSO absorption lines}
\author{C. Fechner \inst{1}}
\institute{Institut f\"{u}r Physik und Astronomie, Universit\"{a}t Potsdam,
  Haus 28, Karl-Liebknecht-Str. 24/25, 14476 Potsdam, Germany\\
  \email{cfech@astro.physik.uni-potsdam.de}}
\offprints{C.\ Fechner,\\ \email{cfech@astro.physik.uni-potsdam.de}}
\date{Received 14 April 2011 / Accepted 06 June 2011}

\abstract{
We present a new approach to constrain the spectral energy distribution of the intergalactic UV background observationally by studying metal absorption systems.

We study single-component metal line systems exhibiting various well-measured species.
Among the observed transitions at least two ratios of ionization stages from the same element are required, e.g.\ \ion{C}{iii}/\ion{C}{iv} and \ion{Si}{iii}/\ion{Si}{iv}.
For each system photoionization models are constructed varying the spectrum of the ionizing radiation.
The spectral energy distribution can then be constrained by comparing the models with the observed column density ratios.

Extensive tests with artificial absorbers show that the spectrum of the ionizing radiation cannot be reconstructed unambiguously, but it is possible to constrain the main characteristics of the spectrum.
Furthermore, the resulting physical parameters of the absorber, such as ionization parameter, metallicity, and relative abundances, may depend strongly on the adopted ionizing spectrum.
Even in case of well-fitting models the uncertainties can be as high as $\sim 0.5\,\mathrm{dex}$ for the ionization parameter and up to $\sim 1.5\,\mathrm{dex}$ for the metallicity.
Therefore, it is essential to know the hardness of the UV background when estimating the metallicity of the intergalactic medium.
Applying the procedure to a small sample of 3 observed single-component metal line systems yields a soft ionizing radiation at $z > 2$ and a slightly harder spectrum at $z < 2$.
The resulting energy distributions exhibit strong \ion{He}{ii} Ly$\alpha$ re-emission features suggesting that reprocessing by intergalactic \ion{He}{ii} is important.
Comparing to UV background spectra from the literature indicates that the recent model of Madau \& Haardt (2009) including sawtooth modulation due to reprocessing by intergalactic \ion{He}{ii} with delayed helium reionization fits the investigated systems very well.

\keywords{quasars: absorption lines -- intergalactic medium -- cosmology: observations}} 

\maketitle

\section{Introduction}

The intergalactic medium (IGM) observed in QSO absorption spectra at redshifts $z < 6$ is highly ionized.
At redshifts $z \gtrsim 1$ photoionization is the dominant ionization process.
To interpret quasar absorption spectra and to investigate the physics of the IGM ionization corrections have to be applied and a thorough understanding of the ionizing radiation field is required.
Related key questions are the formation of galaxies and the enrichment of the IGM with heavy elements.
For studies of the enrichment history the spectrum and evolution of the UV background have to be known to estimate the metallicity of intergalactic absorbers.
It is still under debate how and when metals are distributed into the IGM \citep[e.g.][]{porciani+2005, songaila2006}.
However, the recently detected lack of \ion{C}{iv} absorption systems at $z\sim 6$ suggests that the IGM has not been pre-enriched to a high level \citep{becker+2009, ryanweber+2009}.
Galaxy formation is also influenced since the cooling of the gas is affected by the metallicity \citep{wiersma+2009} as well as by the intensity and spectrum of the ionizing background.
For instance, photoheating suppresses star formation in dwarf galaxies \citep[e.g.\ \citet{efstathiou1992, hoeft+2006}, but see also][]{gnedin2010}.

The highly photoionized stage of the intergalactic material is maintained by the UV background which is produced by the radiation of quasars and galaxies filtered by absorption and re-emission processes in the IGM.
Its spectral energy distribution (SED) can be estimated from theoretical calculations based on the redshift-dependent properties of the sources and the absorbers (e.g.\ the hardness of the QSO spectra, their number density, the column density distribution of intergalactic absorbers, etc.).
The UV background model from \citet[][HM01]{haardt+2001} is usually adopted in absorption line studies.
A particular application is estimating the metallicity of the IGM, which is important to understand the mechanisms and stellar sources of metal enrichment.

Alternative UV background models from other authors \citep{fardal+1998, faucher+2009} and recent upgrading \citep{madau+2009} reveal the same characteristic spectral shape, which is determined by an underlying power law slope of the quasar sources with additional contribution of galaxies at low energies.
Absorption breaks and re-emission features due to the intergalactic \ion{H}{i} and \ion{He}{ii} are superimposed on the spectrum.
The specifications of these characteristics depend on the details of the handling of the cosmological radiative transfer as well as on the adopted properties of the quasars and galaxies as the sources of radiation.
The spectrum in the energy range $\sim$ 1-10\,Ryd is in particular important since the species easily accessible in the optical have ionization potentials in this range (e.g.\ \ion{C}{iii} 3.5\,Ryd, \ion{C}{iv} 4.74\,Ryd, \ion{Si}{iii} 2.46\,Ryd, \ion{Si}{iv} 3.32\,Ryd, \ion{O}{vi} 10.15\,Ryd), which are, therefore, usually adopted to trace the metallicity and the relative abundances of the IGM.
However, from investigating intergalactic \ion{He}{ii} absorption, which probes the spectral hardness of the ionizing radiation, it is known that the UV background is probably fluctuating \citep[e.g.][]{kriss+2001, shull+2004, fechner+2006b, fechner+2007a, shull+2010}.
These fluctuations in spectral hardness can be attributed to variations of the IGM density, effects of radiative transport, as well as local sources close to the line of sight.
Recently, \citet{bolton+2011} found from numerical simulation, that spatial inhomogeneities of the spectral hardness of the UV background have only little impact on the ionization balances of \ion{Si}{iv} and \ion{C}{iv}.
This is due to the longer mean free path of photons with energies close to the \ion{He}{ii} ionization edge.  
However, studies of individual intergalactic metal line systems indicate deviations from the standard HM01 spectrum, in particular in the important energy range 3-4\,Ryd \citep[e.g.][]{agafonova+2005, agafonova+2007, fechner+2006a, reimers+2006}.

In order to investigate in detail the spectral energy distribution of the UV background metal line systems have to be studied by means of photoionization modeling.
The ionization fractions within an absorber will depend on the spectrum and intensity of the ionizing radiation as well as on the density of the absorbing material.
Therefore, the ionizing spectrum can be reconstructed to a given level of confidence for appropriate systems.
Suitable systems have to exhibit various different species, preferably at least two of each observed element to avoid degeneracy with the elemental abundance pattern.
In principle, this can be done using absorption features of \ion{C}{ii}, \ion{C}{iii}, \ion{C}{iv}, \ion{Si}{ii}, \ion{Si}{iii}, \ion{Si}{iv}, and possibly several low-ionization species at redshift $z \sim 2$. 

In this paper we introduce a new approach to constrain the SED of the UV background by extensive photoionization modeling of metal line systems.
The criteria applied to select appropriate absorption systems are explained in Sect.\ \ref{systems}.
Our new method is presented in Sect.\ \ref{method}.
Its potential and limitations are illustrated in Sect.\ \ref{testcases} by applying it to artificially constructed test absorbers.
The selected observed absorption systems are investigated in Sect.\ \ref{results} and the results are discussed and compared to the most recent model calculations in Sect.\ \ref{discussion}.
We summarize our conclusions in Sect.\ \ref{conclusions}.
Throughout the paper abundances are given in the notation $\mathrm{[X/Y]} = \log (\mathrm{X/Y}) - \log (\mathrm{X/Y})_{\sun}$ with solar abundances taken from \citet{asplund+2005}.

\section{Selection of appropriate systems}\label{systems}

\subsection{Selection criteria}

In order to explore the potential of a new approach to constrain the spectral shape of the UV background by modeling metal line systems it is important to select appropriate absorption systems.
Suitable systems should be as simple as possible to avoid systematics due to simplifying assumptions about the physical properties of the absorber.
Therefore, we select systems that (1) show absorption in one single component only, i.e.\ without any substructure, (2) are unblended and unsaturated in all important transitions, and (3) exhibit various species with at least two ratios of two different ions from the same element.
These selection criteria are supposed to reduce implicit assumptions about the characteristics of the absorbing material like velocity structure, multiple gas phases and elemental abundances.

(1) single-component systems: 
Systems exhibiting multiple component metal line features usually show strong \ion{H}{i} absorption which has to be distributed to the individual components.
In principle, the \ion{H}{i} can be assigned to the individual components assuming the same metallicity throughout the system.
However, analyses of damped Ly$\alpha$ (DLA) systems probed by gravitationally lensed sight lines showed that the metallicity can vary on small scales \citep[e.g.][]{lopez+1999, lopez+2003, richter+2005}.
Selecting single-component systems avoids any assumptions about the velocity structure and the distribution of metals within the absorbing gas

(2) unblended and unsaturated absorption features: 
Photoionization models are based on the column densities estimated from the observed absorption profiles.
In particular, the ionization parameter $U = n_{\gamma}/n_{\element{H}}$, which is the ratio of the ionizing photon to hydrogen density, is fixed by a measured column density ratio.
To keep the uncertainties of the model as low as possible it is important to have exact column density estimates.
Moreover, we will consider the column density measurements to evaluate the tested SEDs, and the constraints will be more strict if the error bars of the column densities are small.
Line parameters can be measured best if the absorption features are neither saturated nor blended.

(3) various species with at least two ratios of two different ions from the same element:
the ionization parameter of an observed system can be constrained by comparing a measured column density ratio to predictions from photoionization calculations adopting a radiation field.
For a given normalization of the ionizing spectrum the ionization parameter is uniquely related to the density $n_{\element{H}}$ of the absorber.
In principle, for any spectral energy distribution a solution can be found, even though extremely hard/soft spectra will yield unrealistically low/high densities and probably unreasonable metallicity values.
Thus, the derived physical parameters rely on a reasonable choice of the radiation field.
If two or more column density ratios are available, both have to be reproduced at the same ionization parameter.
This will be impossible for unrealistic ionizing spectra providing a tool to evaluate the plausibility of various SEDs.
In order to be independent of assumptions about the relative abundances, which may deviate from solar in the IGM \citep[e.g.][]{aguirre+2004, aguirre+2008, fechner+2009}, the adopted ratios have to consider two species of the same element.
For the systems that will be studied in the following the ratios \ion{C}{iii}/\ion{C}{iv} and \ion{Si}{iii}/\ion{Si}{iv} or \ion{Si}{ii}/\ion{Si}{iii}/\ion{Si}{iv} and \ion{Al}{ii}/\ion{Al}{iii}, respectively, are available.

We have searched in published data for metal absorption systems satisfying these selection criteria.
Three systems at different redshifts are selected which are suitable for a feasibility study.

\begin{table}
  \caption[]{Fitted line parameters.}
  \label{par_systems}
  $$
  \begin{array}{l l c c r}
    \hline\hline
    \noalign{\smallskip}
\mathrm{QSO} &\mathrm{ion} & z & \log N & b (\mathrm{km\,s}^{-1})\\     
    \noalign{\smallskip}
    \hline
    \noalign{\smallskip}
\mathrm{HE~1347-2457}&\ion{H}{i}   & 1.7529056 & 15.486\pm0.112 &17.56\pm0.50\\
                     &\ion{C}{ii}  & 1.7528981 & 14.087\pm0.016 & 6.03\pm0.10\\
                     &\ion{C}{iv}  & 1.7528915 & 14.620\pm0.018 & 7.46\pm0.06\\
                     &\ion{Mg}{ii} & 1.7528971 & 12.816\pm0.008 & 4.50\pm0.06\\
                     &\ion{Al}{ii} & 1.7529035 & 11.600\pm0.013 & 6.09\pm0.31\\
                     &\ion{Al}{iii}& 1.7529033 & 12.229\pm0.006 & 4.64\pm0.12\\
                     &\ion{Si}{ii} & 1.7529005 & 12.910\pm0.018 & 4.79\pm0.11\\
                     &\ion{Si}{iii}& 1.7529052 & 13.553\pm0.142 & 6.30\pm0.42\\
                     &\ion{Si}{iv} & 1.7528871 & 13.621\pm0.007 & 6.29\pm0.05\\
    \noalign{\smallskip}
    \hline
    \noalign{\smallskip}
\mathrm{HS~1700+6416}&\ion{H}{i}   & 2.3798685 & 15.414\pm0.170 &26.49\pm1.55\\
                     &\ion{C}{iii} & 2.3799345 & 13.495\pm0.187 & 8.64\pm1.63\\
                     &\ion{C}{iv}  & 2.3799374 & 13.059\pm0.005 & 9.69\pm0.14\\
                     &\ion{Si}{iii}& 2.3799212 & 11.689\pm0.035 & 6.66\pm0.84\\
                     &\ion{Si}{iv} & 2.3799238 & 11.689\pm0.037 & 5.74\pm0.99\\
    \noalign{\smallskip}
    \hline
    \noalign{\smallskip}
\mathrm{HE~0940-1050}&\ion{H}{i}   & 2.8265295 & 14.612\pm0.005 &20.13\pm0.22\\
                     &\ion{C}{iii}^{\,\mathrm{a}} & 2.8265585 & 13.232\pm0.013 & 6.35\pm0.00\\
                     &\ion{C}{iv}  & 2.8265585 & 13.218\pm0.005 & 6.35\pm0.13\\
                     &\ion{Si}{iii}^{\,\mathrm{a}}& 2.8265616 & 11.461\pm0.018 & 6.59\pm0.00\\
                     &\ion{Si}{iv} & 2.8265616 & 12.037\pm0.020 & 6.59\pm0.02\\
    \noalign{\smallskip}
    \hline
  \end{array}
  $$
\begin{list}{}{}
  \item[$^{\mathrm{a}}$] redshifts and Doppler-parameters of \ion{C}{iii} and \ion{Si}{iii} have been set to the measured values of the \ion{C}{iv} and \ion{Si}{iv} profiles, respectively
\end{list}
\end{table}

\subsection{System at $z = 1.7529$ towards HE~1347-2457}

\begin{figure}
  \centering
  \resizebox{\hsize}{!}{\includegraphics[bb=35 325 350 780,clip=]{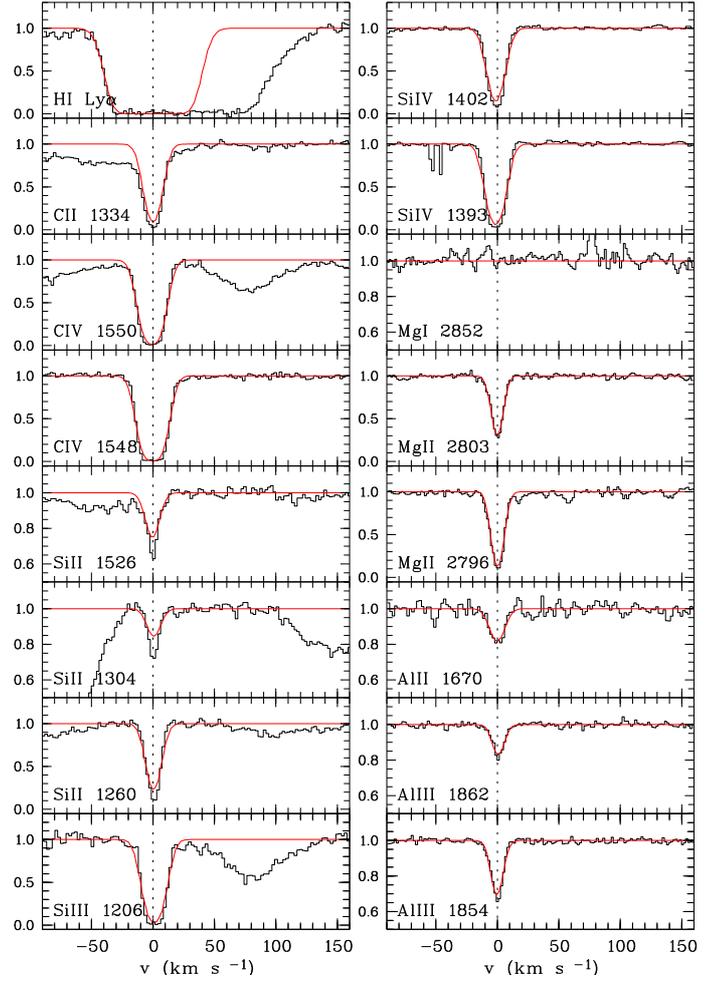}}
  \caption{Profiles of the systems at $z = 1.7529$ towards HE~1347-2457. 
The histogram-like lines represent the observed data.
The solid lines indicate the profiles from the line fit.
  }
  \label{profiles_z1.7529}
\end{figure}

HE~1347-2457 ($z_{\mathrm{em}} = 2.61$) has been observed as part of the ESO Large Program "Cosmic evolution of the IGM" \citep{bergeron+2004} with UVES at the VLT.
The optical spectrum has a resolution $R \approx 45\,000$ and a signal-to-noise ratio of $S/N \sim 35 -70$.
For details of the data reduction we refer to \citet{aracil+2004}.

The system at $z = 1.7529$ exhibits one single absorption component predominately in low-ionization species.
Features of \ion{C}{ii}, \ion{Mg}{ii}, \ion{Al}{ii}, \ion{Al}{iii}, \ion{Si}{ii}, \ion{Si}{iii}, \ion{Si}{iv}, and also \ion{C}{iv} are identified.
Column densities are estimated by fitting Doppler profiles.
The resulting line parameters are listed in the upper part of Table~\ref{par_systems} and the observed features together with the modeled profiles are presented in Fig.\ \ref{profiles_z1.7529}.

Unfortunately, \ion{H}{i} is only detected in Ly$\alpha$ and the corresponding feature is saturated. 
Therefore, the column density estimate may be wrong leading to an incorrect metallicity.
However, since the absorption is optically thin (i.e.\ $N_{\ion{H}{i}} \lesssim 10^{17}\,\mathrm{cm}^{-2}$), the modeled column density ratios of the metal ions do not depend on the exact value of the \ion{H}{i} column density.
Thus, our main conclusions will not be affected by this uncertainty.
However, we should keep it in mind for the interpretation of derived metallicity.
The features of \ion{C}{iv} and \ion{Si}{iii} as well as \ion{Si}{iv} are mildly saturated. 
But since the estimated Doppler-parameters of these species agree well, we have confidence in the column density measurements.

\subsection{System at $z = 2.3799$ towards HS~1700+6416}

\begin{figure}
  \centering
  \resizebox{\hsize}{!}{\includegraphics[bb=35 485 350 780,clip=]{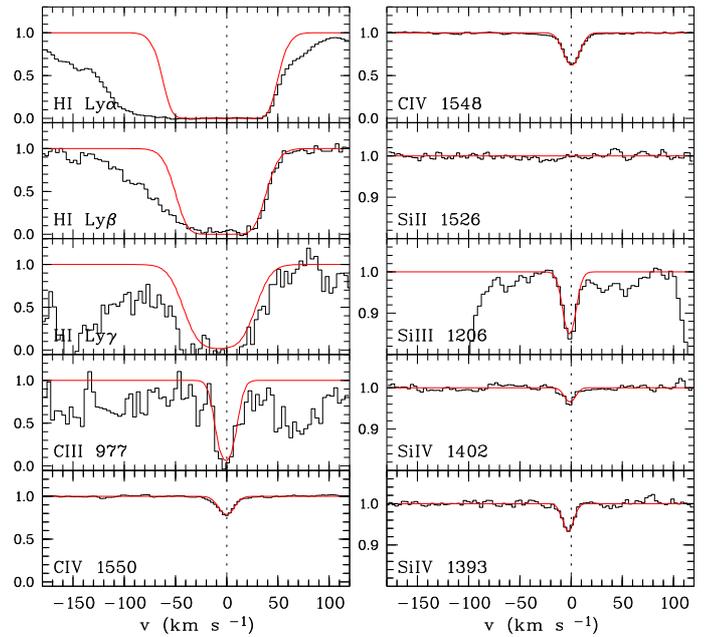}}
  \caption{Profiles of the systems at $z = 2.3799$ towards HS~1700+6416.
The histogram-like lines represent the observed data.
The solid lines indicate the profiles from the line fit.
  }
  \label{profiles_z2.3799}
\end{figure}

The metal line systems in the line of sight towards the QSO HS~1700+6416 ($z_{\mathrm{em}} = 2.72$) have been studied by various authors \citep[e.g.][]{vogel+1995, koehler+1996, petitjean+1996, tripp+1997, simcoe+2006, fechner+2006a}.
We use the co-added spectrum described in \citet{fechner+2006a} of the data presented in \citet{songaila1998} and \citet{simcoe+2002}.
The resolution is $R \sim 38\,000$ with $S/N \sim 100$ at $4000\,\mathrm{\AA}$.

The selected system at $z = 2.3799$ exhibits features of \ion{H}{i}, \ion{C}{iii}, \ion{C}{iv}, \ion{Si}{iii}, and \ion{Si}{iv}.
The line parameters derived by Doppler-profile fitting are summarized in the middle part of Table \ref{par_systems} and the observed features are presented in Fig.\ \ref{profiles_z2.3799}.
Again the \ion{H}{i} absorption is saturated.
The column density estimate is based on a simultaneous fit of all available transitions of the Lyman series (Ly$\alpha$, $\beta$, $\gamma$).
Features of \ion{Si}{iii} and \ion{Si}{iv} are weak but clearly detected.
The \ion{C}{iii} profile is slightly saturated and the corresponding column density estimate is the most uncertain one in this system.

\subsection{System at $z = 2.8266$ towards HE~0940-1050}

\begin{figure}
  \centering
   \resizebox{\hsize}{!}{\includegraphics[bb=35 485 350 780,clip=]{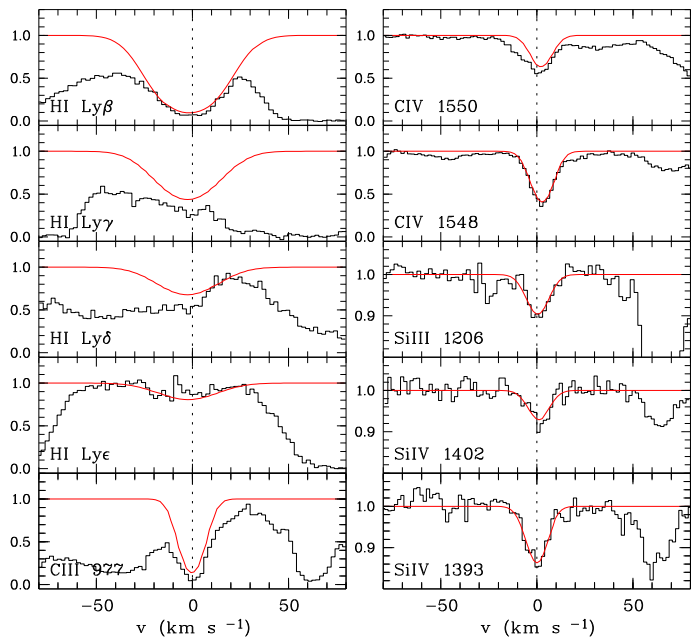}}
  \caption{Profiles of the systems at $z = 2.8266$ towards HE~0940-1050.
The histogram-like lines represent the observed data.
The solid lines indicate the profiles from the line fit.
  }
  \label{profiles_z2.8266}
\end{figure}

As HE~1347-2457 the QSO HE~0940-1050 ($z_{\mathrm{em}} = 3.08$) is part of the ESO Large Program sample.
Data quality and reduction process are equivalent for both lines of sight.

We select a system at $z = 2.8266$ showing absorption in \ion{C}{iii}, \ion{C}{iv}, \ion{Si}{iii}, and \ion{Si}{iv}.
Again, the features of \ion{Si}{iii} and \ion{Si}{iv} are weak and the \ion{C}{iii} estimate is the most uncertain due to interlopers from the Ly$\alpha$ forest.
Redshifts and $b$-parameters of \ion{C}{iii} and \ion{Si}{iii} have been fixed to the values derived for \ion{C}{iv} and \ion{Si}{iii}, respectively, to reduce the uncertainties in the column density estimates.
Using the Lyman series up to Ly$\epsilon$ the column density of \ion{H}{i} is well-measured.
We summarize the derived line parameters in the lower part of Table \ref{par_systems} and present the observed profiles in Fig.\ \ref{profiles_z2.8266}.

\section{Modeling}\label{method}

\begin{figure}
  \centering
   \resizebox{\hsize}{!}{\includegraphics[bb=35 450 485 775,clip=]{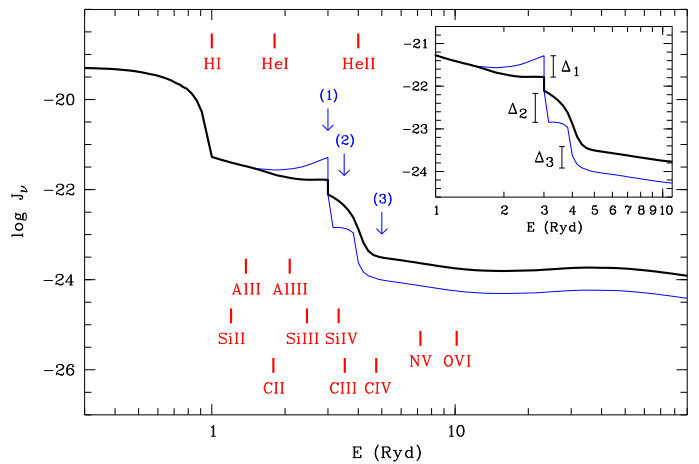}}
  \caption{Spectral energy distribution of the ionizing radiation in comparison to the ionization potential of several species.
The thick line represents the UV background at $z \sim 2.0$ according to the model of \citep[][HM01]{haardt+2001}, which is adopted as ionizing radiation for test case A (see Sect.\ \ref{testcases}).
The thin line indicates an example of a modified spectrum, where the main characteristics varied in this work are marked with arrows.
They are measured in deviation from the HM01 spectrum in dex as indicated in the inset.
These are (1) the height of the \ion{He}{ii} Ly$\alpha$ re-emission peak at 3\,Ryd $\Delta_1$, (2) the flux level in the energy range $3 < E < 4\,\mathrm{Ryd}$ $\Delta_2$, and (3) the depth of the \ion{He}{ii} break at 4\,Ryd $\Delta_3$.
The presented modified SED can parameterized with $\Delta_1 = +0.5$, $\Delta_2 = -0.7$, and $\Delta_3 = -0.5$, and is adopted for test case B in Sect.\ \ref{testcases}.
  }
  \label{uvb}
\end{figure}

Photoionization models are computed for each system using CLOUDY \citep[v05.07.06;][]{ferland+1998}.
The basic idea is to construct individual best-fitting models for several spectra of the ionizing radiation.
Since the systems are selected to provide two independent column density ratios, the tested SEDs can then be evaluated regarding their consistency with the observations.
Furthermore, the sensitivity of the resulting physical parameters of the absorbers, e.g.\ density and metallicity, on the spectrum of the ionizing radiation can be investigated.

For a given radiation field the ionization parameter $U$ is determined by matching the observed column density ratio of two species.
Because of the uncertainties of the \ion{C}{iii} measurement we adopt \ion{Si}{iii}/\ion{Si}{iv} for all systems.
The metallicity of the absorber is then adjusted to recover the absolute values of the measured column densities.
We assume a solar abundance pattern and reference solar abundances are taken from \citet{asplund+2005}.
The tested SEDs are scaled to $J_{\ion{H}{i}} = 10^{-21.1}\,\mathrm{erg\,s}^{-1}\mathrm{cm}^{-2}\mathrm{Hz}^{-1}\mathrm{sr}^{-1}$ at 1\,Ryd \citep{scott+2000}.
With this normalization the number of ionizing photons of each SED is fixed and the derived ionization parameter corresponds to a density $n_{\element{H}}$.
Thus, for an individual SED a best-fit model is found determined by the resulting values of $\log n_{\element{H}}$ and $\log U$, respectively, the metallicity [Si/H], and the relative abundance [Si/C], which is scaled a posteriori to match as well the measured column densities of the carbon species as accurate as possible.

This procedure is applied to various ionizing spectra.
At first we adopt the UV background from \citet[][HM01]{haardt+2001} at the appropriate redshift.
Its SED at $z \sim 2.0$ is indicated as thick line in Fig.\ \ref{uvb}.
The shape of this spectrum represents a mean UV background constituted by the radiation of galaxies and quasars which is filtered when propagating through the IGM.
The main characteristics are a break at 1\,Ryd due to absorption by intergalactic \ion{H}{i}, a second break at 4\,Ryd due to absorption by intergalactic \ion{He}{ii}, and a small peak at 3\,Ryd due to \ion{He}{ii} Ly$\alpha$ re-emission.
The galaxies included are assumed to have an escape fraction of ionizing photons of 10\%.
Their soft radiation mainly contributes to the SED at energies $\lesssim 1\,\mathrm{Ryd}$.

Inspired by the results of \citet{fechner+2006a} and \citet{agafonova+2007}, three characteristics of the HM01 spectrum are systematically varied: (1) the height of the 3\,Ryd Ly$\alpha$ peak, (2) the depletion of the intensity in the range $3 < E < 4\,\mathrm{Ryd}$, and (3) the strength of the 4\,Ryd break.
We measure each of the modifications in departure from the HM01 spectrum in dex denoted by $\Delta_1$, $\Delta_2$, and $\Delta_3$ in the following, as indicated in the inset in Fig.\ \ref{uvb}.
The thin line in Fig.\ \ref{uvb} shows an example of a modified SED with an enhanced 3\,Ryd Ly$\alpha$ peak by $\Delta_1 = 0.5$, intensity depletion at 3-4\,Ryd by $\Delta_2 = -0.7$, and a deeper break at $>4$\,Ryd by $\Delta_3 = -0.5$.

The parameter space explored is $-0.4 \le \Delta_1 \le +1.0$ using a step size of 0.1\,dex, i.e.\ we vary between spectra with no 3\,Ryd Ly$\alpha$ peak at all (which is at $\Delta_1 = -0.4$ at the probed redshifts) to a pronounced peak 1.0\,dex stronger than in the HM01 spectrum.
The intensity between 3 and 4\,Ryd is reduced by upto 1.0\,dex ($-1.0 \le \Delta_2 \le 0.0$) with a step size of 0.1\,dex.
The depth of the 4\,Ryd break is modified in the range $-1.0 \le \Delta_3 \le +0.4$ with a step size of 0.2\,dex.
This means that the intensity at $E > 4\,\mathrm{Ryd}$ is either reduced by upto 1\,dex or increased by upto 0.4\,dex, where the latter modification generates a substantially harder SED.
Since all modifications are combined with each other in this feasibility study, 1320 models are computed in total for each system.

The resulting models are evaluated to investigate which of the tested SEDs are consistent with the observed metal line systems to a given level of confidence.
Therefore, we use the second observed column density ratio.
In case of the systems at $z > 2$ this is \ion{C}{iii}/\ion{C}{iv}.
Comparing the modeled ratio to the observed value $\log(N_1/N_2) \pm \Delta\log(N_1/N_2)$, where $\Delta\log(N_1/N_2)$ denotes the $1\,\sigma$ error of the column density ratio, the confidence of the best-fit models are estimated.
In case of the system at $z = 1.7529$ two more ratios are available (\ion{Si}{ii}/\ion{Si}{iii} and \ion{Al}{ii}/\ion{Al}{iii}) which are applied individually and in a combined manner, respectively, as we will present in Sect.\ \ref{results}.
Evaluation of the additional column density ratio then leads to the SED producing the overall best-fitting model.
Furthermore, we can derive confidence levels for the parameters of the SED modifications.

\section{Test cases}\label{testcases}

\begin{table}
  \caption[]{Input parameters and column densities of the test cases.}
  \label{par_testcase}
  $$
  \begin{array}{c c c c c c c c c}
    \hline\hline
    \noalign{\smallskip}
\mathrm{case} &\log n_{\element{H}}&\mathrm{[Si/H]}&\mathrm{[Si/C]}&\ion{H}{i}&\ion{C}{iii}&\ion{C}{iv}&\ion{Si}{iii}&\ion{Si}{iv}\\
    \noalign{\smallskip}
    \hline
    \noalign{\smallskip}
\mathrm{A}^{\,\mathrm{a}} & -3.5 & -1.8 & +0.7 & 15.30 & 12.659 & 12.967 & 11.133 & 11.512 \\
\mathrm{B}^{\,\mathrm{b}} & -3.5 & -1.8 & +0.7 & 15.30 & 13.402 & 13.215 & 11.949 & 12.516 \\
    \noalign{\smallskip}
    \hline
  \end{array}
  $$
  \begin{list}{}{}
    \item[$^{\mathrm{a}}$] absorber ionized by an original HM01 spectrum at $z \sim 2$ represented by the thick line in Fig.\ \ref{uvb}
    \item[$^{\mathrm{b}}$] absorber ionized by a modified HM01 spectrum with $\Delta_1 = +0.5$, $\Delta_2 = -0.7$, and $\Delta_3 = -0.5$ represented by the thin line in Fig.\ \ref{uvb}
  \end{list}
\end{table}

In order to test the ability of the procedure described above to recover the spectrum of the ionizing radiation we construct two artificial systems.
Presuming an absorber with $\log n_{\element{H}} = -3.5$, $[\mathrm{Si/H}] = -1.8$, $[\mathrm{Si/C}] = +0.7$, and an observed \ion{H}{i} column density of $\log N_{\ion{H}{i}} = 15.3$ is exposed to an unmodified HM01 spectrum at $z \sim 2$ (corresponding to the thick line in Fig.\ \ref{uvb}), CLOUDY predicts column densities for \ion{C}{iii}, \ion{C}{iv}, \ion{Si}{iii}, and \ion{Si}{iv} listed in Table \ref{par_testcase} as case A.
As a second test case B the same absorber is exposed to a modified HM01 spectrum with $\Delta_1 = +0.5$, $\Delta_2 = -0.7$, and $\Delta_3 = -0.5$, indicated by the thin line in Fig.\ \ref{uvb}.
For both test cases we run the same grid of models to establish whether the presumed ionizing spectra can be recovered.

In the following the best-fit photoionization models are evaluated considering different aspects about the robustness of the method and the influence of the individual parameters on the results.
To find the best-fit model for every individual SED the \ion{Si}{iii}/\ion{Si}{iv} ratio is adopted.
The \ion{C}{iii}/\ion{C}{iv} ratio is then used additionally to evaluate the goodness of the model with respect to the adopted SED.
Assuming an uncertainty of $\Delta\log N = 0.01$ for each column densities we adopt a $1\,\sigma$ error of $\Delta\log(N_\ion{C}{iii}/N_\ion{C}{iv}) = \sqrt{(\Delta\log N_{\ion{C}{iii}})^2+(\Delta\log N_{\ion{C}{iv}})^2} = 0.014$ for the \ion{C}{iii}/\ion{C}{iv} ratio.
Thus, models yielding $|(\ion{C}{iii}/\ion{C}{iv})_{\mathrm{obs}}-(\ion{C}{iii}/\ion{C}{iv})_{\mathrm{mod}}| \le 0.014$ reproduce the column densities within $1\,\sigma$ confidence.
Here, $(\ion{C}{iii}/\ion{C}{iv})_{\mathrm{obs}}$ denotes the column density ratio of the test absorbers given in Table \ref{par_testcase}.

\subsection{Recovery of the ionizing spectrum}

\begin{figure*}
  \centering
   \resizebox{\hsize}{!}{\includegraphics[bb=0 450 585 685,clip=]{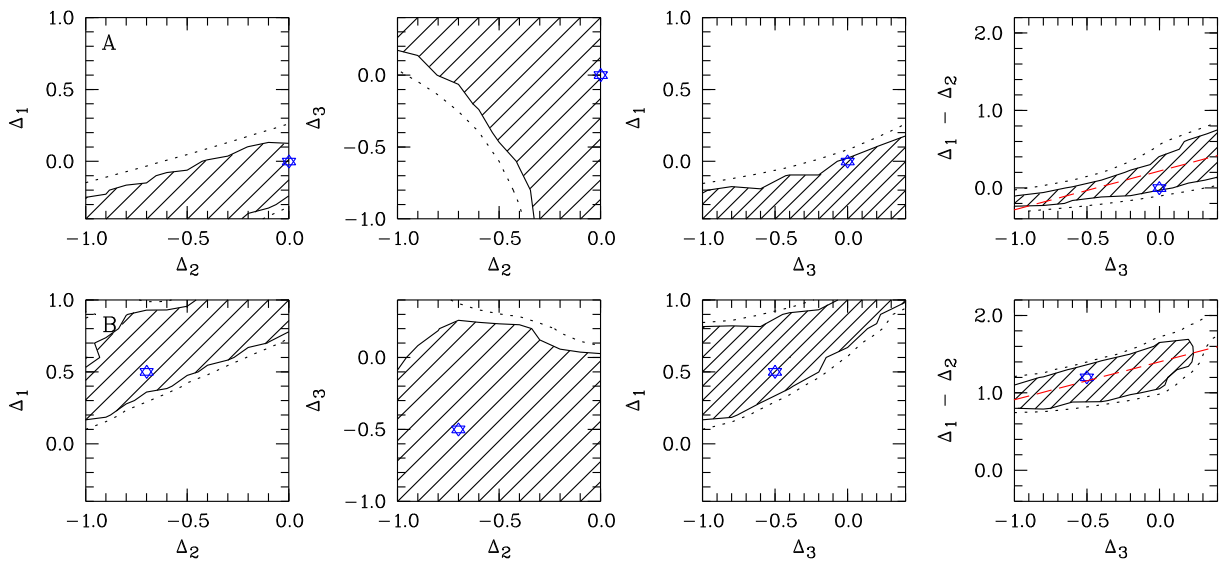}}
  \caption{$1\,\sigma$ (marked areas) and $3\,\sigma$ (dotted lines) confidence regions of the tested SED parameters for test cases A (upper panels) and B (lower panels).
The three left panels show the projections into each of the three possible planes, i.e.\ $\Delta_1$ versus $\Delta_2$, $\Delta_3$ versus $\Delta_2$, and $\Delta_1$ versus $\Delta_3$, respectively.
The right panels show the enhancement of the intensity break at 3\,Ryd compared to the HM01 background, which means $\Delta_1 - \Delta_2$, versus the strength of the 4\,Ryd break $\Delta_3$, where the dashed line indicates a linear fit of the correlation (see text).
The parameters of the input SEDs are marked with stars.
  }
  \label{projection_test}
\end{figure*}

\begin{figure}
  \centering
   \resizebox{\hsize}{!}{\includegraphics[bb=30 510 330 695,clip=]{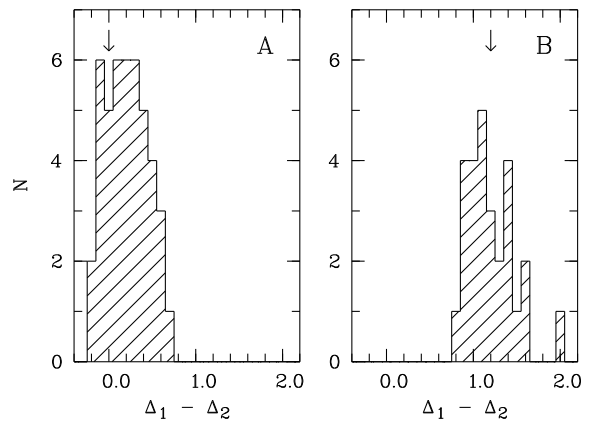}}
  \caption{Distribution of $\Delta_1 - \Delta_2$ for UV background spectra that fit the observed column density ratios up to $1\,\sigma$ confidence for test case A (left panel) and B (right panel).
    Input values are marked with an arrow.
  }
  \label{delta_test}
\end{figure}

In Fig.\ \ref{projection_test} the $1\,\sigma$ and $3\,\sigma$ confidence regions of the SED parameters are shown for test case A (upper panels) and B (lower panels) projected into each of the three possible planes in parameter space ($\Delta_1$ versus $\Delta_2$, $\Delta_3$ versus $\Delta_2$, and $\Delta_1$ versus $\Delta_3$; from left to right).
Even though the parameters of the input SEDs (marked with stars) are within the $1\,\sigma$ confidence regions, they are not recovered unambiguously.

From the left-most panel it can be seen that the favored $\Delta_1$ and $\Delta_2$ values seem to be correlated.
This can be understood since both, the height of the 3\,Ryd Ly$\alpha$ peak and the intensity reduction in the range 3-4\,Ryd are related to the \ion{He}{ii} recombination re-emission.
Since the models are based on the \ion{Si}{iii}/\ion{Si}{iv} ratio, they are in particular sensitive to the spectral shape at 3\,Ryd since the ionization potentials of \ion{Si}{iii} ($2.46\,\mathrm{Ryd}$) and \ion{Si}{iv} ($3.32\,\mathrm{Ryd}$) are located in this energy range.
Thus, the important characteristic is the strength of the intensity break at 3\,Ryd rather than the absolute values of the intensity just below and above this energy.
Therefore, we introduce the strength of the break a 3\,Ryd relative to the HM01 spectrum as an alternative parameter, which is $\Delta_1 - \Delta_2$.
The distribution of $\Delta_1 -\Delta_2$ for the $1\,\sigma$-confidence models are shown in Fig.\ \ref{delta_test}.
For both test cases a clear maximum is seen.
For test case A we recover $\langle\Delta_1-\Delta_2\rangle = 0.2\pm^{0.5}_{0.4}$ which slightly overestimates the actual value (which is $0.0$ since the ionizing radiation is the unmodified HM01 spectrum).
Test case B has $\langle\Delta_1 - \Delta_2\rangle = 1.2\pm^{0.8}_{0.4}$, which is recovered by the procedure even though with larger scatter.

The confidence contours for this parameter versus the strength of the intensity break at 4\,Ryd $\Delta_3$ are shown in the right panels of Fig.\ \ref{projection_test}.
The 3\,Ryd break $\Delta_1 - \Delta_2$ appears to be weakly correlated with $\Delta_3$ in the sense that larger $\Delta_3$, i.e.\ harder spectra, favor stronger breaks at 3\,Ryd.
Since the 4\,Ryd break is the \ion{He}{ii} ionization edge, it is governed as well by the amount of \ion{He}{ii} in the IGM.
Therefore, a correlation between $\Delta_1 - \Delta_2$ and $\Delta_3$ is expected, since more \ion{He}{ii} means both: stronger \ion{He}{ii} Lyman continuum absorption and therefore a stronger 4\,Ryd break, as well as stronger \ion{He}{ii} recombination re-emission and stronger \ion{He}{ii} Lyman series absorption (see Fig.\ 1 of \citet{madau+2009} for illustration).
Describing this correlation roughly as linear we fit $\Delta_1-\Delta_2 = (0.50\pm0.05)\cdot\Delta_3+(0.22\pm0.02)$ for case A and $(0.49\pm0.10)\cdot\Delta_3+(1.40\pm0.05)$ for case B, indicated as dashed lines in the right panels of Fig.\ \ref{projection_test}.
In general, the adopted method does not lead to reasonable constraints of the intensity level at $E > 4\,\mathrm{Ryd}$.
Models with $1\,\sigma$ confidence can be found for every value of $\Delta_3$ for case A and only the hardest SEDs are excluded by the $1\,\sigma$ constraint for case B.
However, a rough estimate of $\Delta_3$ by inserting the recovered $\Delta_1-\Delta_2$ into the fitted relation yields $\Delta_3 \simeq 0.0$ and $\Delta_3 \simeq -0.4$ for case A and B, respectively, which recovers the input value within 0.1\,dex.
Fixing $\Delta_3$ to the derived value, the best-fit models are then found at $(\Delta_1, \Delta_2, \Delta_3) \simeq (0.0, 0.0, 0.0)$ for case A and $(+0.4, -1.0, -0.4)$ for case B, respectively.
While the HM01 spectrum is perfectly reproduced, the main characteristics of the modified SED (enhanced 3\,Ryd Ly$\alpha$ peak, depression of the intensity at energies $> 3\,\mathrm{Ryd}$ and $> 4\,\mathrm{Ryd}$) are also recovered.
Therefore, is should at least be possible to give reliable constraints on the spectral shape of the UV background also for observed metal line systems.
We expect, however, to find various SEDs that fit the observed column densities to the same level of confidence.
Furthermore, we should keep in mind, that the parameters of the best-fit models are subject to high uncertainties.
Strictly speaking, the parameter ranges marked in Fig.\ \ref{projection_test} indicate the actual error bars of our estimate.

\subsection{Estimated physical parameters}

\begin{figure*}
  \centering
   \resizebox{\hsize}{!}{\includegraphics[bb=48 27 285 518,clip=,angle=-90]{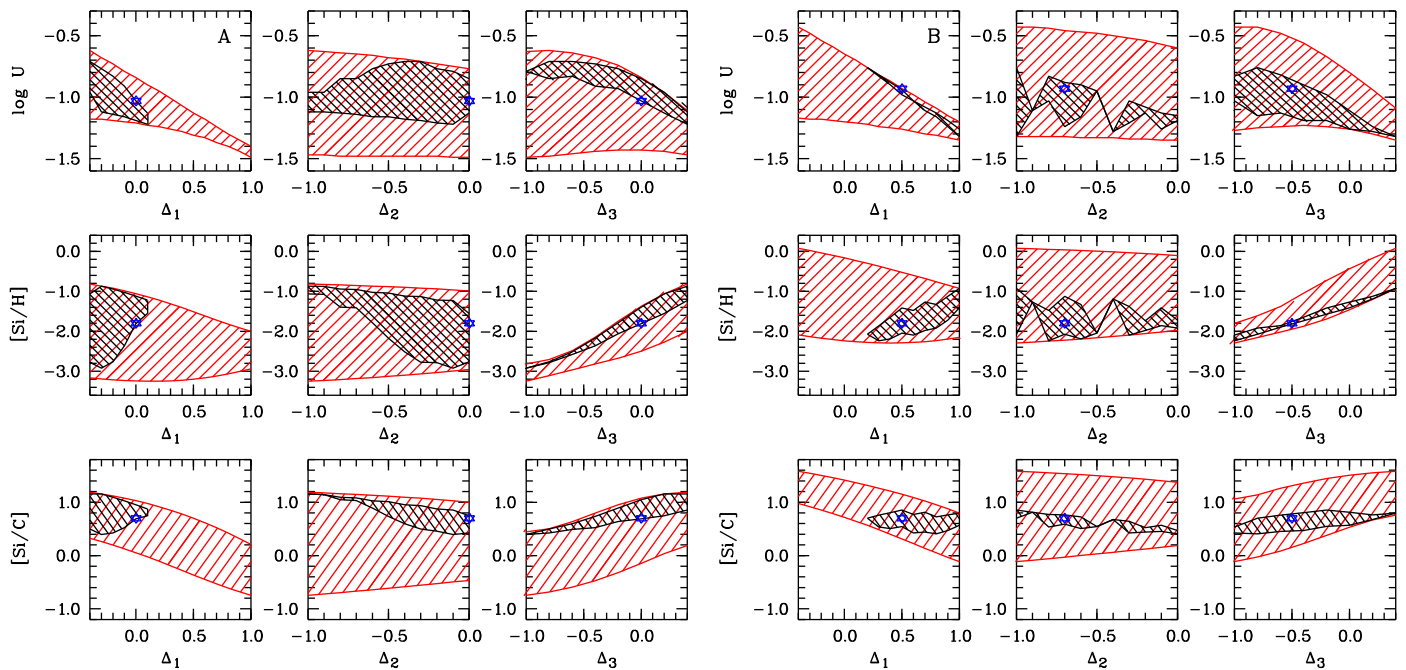}}
  \caption{Recovered physical parameters of the test systems A (left panels) and B (right panels).
Marked areas in the upper, middle, and lower panels show the ionization parameter $\log U$, metallicity [Si/H] and relative abundance [Si/C], respectively, in dependence of the SED parameters $\Delta_1$, $\Delta_2$, and $\Delta_3$ (from the left to the right) estimated for all tested spectra.
The narrower ranges represent the results of the models which are consistent with the "observed" column density ratios on a $1\,\sigma$ level.
The actual input values (see Table \ref{par_testcase}) are indicated as stars.
  }
  \label{parameters_test}
\end{figure*}

\begin{table}
  \caption[]{Recovered physical parameters for the test cases.
}
  \label{par_recover}
  $$
  \begin{array}{c l l l l l}
    \hline\hline
    \noalign{\smallskip}
\mathrm{case} & &\log n_{\element{H}}& \log U  &\mathrm{[Si/H]}&\mathrm{[Si/C]}\\
    \noalign{\smallskip}
    \hline
    \noalign{\smallskip}
\mathrm{A}& \mathrm{input~values} & -3.50 & -1.03 & -1.80 & +0.70\\
& \mathrm{all~models} &-3.35\pm^{0.59}_{0.64}&-1.11\pm^{0.49}_{0.38}&-2.34\pm^{1.52}_{0.91}&+0.32\pm^{0.87}_{1.06}\\
&1\,\sigma &-3.59\pm^{0.32}_{0.30}&-0.97\pm^{0.26}_{0.26}&-1.66\pm^{0.79}_{1.27}&+0.84\pm^{0.32}_{0.44}\\
& 1\,\sigma\,(\Delta_3=\phantom{-}0.0) &-3.65\pm^{0.16}_{0.10}&-0.92\pm^{0.07}_{0.11}&-1.57\pm^{0.17}_{0.24}&+0.90\pm^{0.15}_{0.21}\\
& \mathrm{recovered~SED}^{\,\mathrm{a}}& -3.49 & -1.03 & -1.81 & +0.69\\
    \noalign{\smallskip}
\mathrm{B}& \mathrm{input~values} & -3.50 & -0.93 & -1.80 & +0.70\\
& \mathrm{HM01}& -3.55 & -0.97 & -0.80 & +1.15\\
& \mathrm{all~models}&-3.48\pm^{0.60}_{0.70}&-0.97\pm^{0.54}_{0.38}&-1.33\pm^{1.40}_{0.96}&+0.89\pm^{0.69}_{1.00}\\
&1\,\sigma &-3.32\pm^{0.40}_{0.43}&-1.05\pm^{0.29}_{0.27}&-1.69\pm^{0.76}_{0.55}&+0.63\pm^{0.23}_{0.22}\\
& 1\,\sigma\,(\Delta_3=-0.4) &-3.34\pm^{0.24}_{0.23}&-1.04\pm^{0.14}_{0.15}&-1.70\pm^{0.05}_{0.04}&+0.63\pm^{0.17}_{0.16}\\
& \mathrm{recovered~SED}^{\,\mathrm{b}}& -3.56 & -0.90 & -1.65 & +0.81\\
    \noalign{\smallskip}
    \hline
  \end{array}
  $$
  \begin{list}{}{}
    \item[$^{\mathrm{a}}$] recovered SED is the HM01 spectrum, i.e.\ $(\Delta_1, \Delta_2, \Delta_3) = (0.0, 0.0, 0.0)$
    \item[$^{\mathrm{b}}$] recovered SED has $(\Delta_1, \Delta_2, \Delta_3) = (+0.4, -1.0, -0.4)$
  \end{list}
\end{table}

Usually, photoionization models for metal absorption systems are used to derive physical parameters of the absorbing material.
Most important are the metallicity and the ionization parameter $U$, which is related to the density $n_{\element{H}}$ of the absorber for a given normalization of the ionizing spectrum.
With our grid of best-fit models we are able to investigate the dependence of these physical quantities on the shape of the ionizing radiation.
Since the test cases are based on CLOUDY calculations with given ionizing spectra, it is easily verified that the ionization parameter (and thus the density), the metallicity, and the [Si/C] relative abundance are well reproduced by the CLOUDY models if the correct SED is adopted.
If system B is modeled with a HM01 background as it would be done in a standard analysis, the density would be slightly underestimated (by 0.05\,dex), but the abundances would be heavily overestimated.
The HM01 model yields $\mathrm{[Si/H]} = -0.8$ and $\mathrm{[Si/C]} = +1.2$ which overestimate the actual values by 1.0 and 0.5\,dex, respectively.

Generally, the physical parameters depend on the characteristics of the assumed ionizing spectrum.
In particular, harder SEDs lead to higher metallicities.
The marked areas in Fig.\ \ref{parameters_test} show the ionization parameter $\log U$ (upper panels), the metallicity [Si/H] (middle panels), and the relative abundance [Si/C] (lower panels) resulting from the best-fit model of each tested spectrum versus the SED parameters $\Delta_1$, $\Delta_2$, and $\Delta_3$ (left, middle, right panels), respectively, for test case A (left) and B (right).
The parameters of the models with $1\,\sigma$ confidence are indicated by the narrower marked areas.
Considering all models, harder spectra, i.e.\ larger $\Delta_3$ lead to higher metallicity estimates and higher [Si/C] as expected.
The ionization parameter depends on $\Delta_1$, i.e.\ the height of the 3\,Ryd Ly$\alpha$ peak, in the sense that a more pronounced peak leads to lower $\log U$ and thus to a higher density $\log n_{\element{H}}$.
This is due to using the \ion{Si}{iii}/\ion{Si}{iv} ratio to fix the ionization parameter since the ionization potential of \ion{Si}{iii}, i.e.\ the energy to create \ion{Si}{iv} from \ion{Si}{iii} is 2.46\,Ryd, which is affected by the value of $\Delta_1$ \citep[see Fig.\ \ref{uvb} and also][]{agafonova+2007}.
The estimated [Si/C] abundance decreases with increasing 3\,Ryd Ly$\alpha$ peak as well.
Virtually no correlation is found between the flux level in the range 3 to 4\,Ryd and the derived physical parameters.

Considering all tested spectra the estimated values are spread over a wide range (see also Table \ref{par_recover}).
Therefore, if the explored SEDs represent realistic spectral shapes of the UV background, the uncertainties of the derived physical parameters are in the order of $\gtrsim 1\,\mathrm{dex}$.
The values of the ionization parameter spread over $\Delta\log U \simeq 0.9$, where the corresponding densities cover $\Delta\log n_{\element{H}} \simeq 1.2$.
The abundance estimates are scattered over an even wider range yielding $\Delta\mathrm{[Si/H]} \simeq 2.4$ and $\Delta\mathrm{[Si/C]} \simeq 1.8$ for both test cases.

The range of possible values can be further constrained if only those SEDs are taken into account that fit as well the additional column density ratio \ion{C}{iii}/\ion{C}{iv} within $1\,\sigma$ confidence (dark areas in Fig.\ \ref{parameters_test}).
The mean values are listed in Table \ref{par_recover} where the quoted error bars refer to the range of covered values.
The density recovered for case A is slightly underestimated (by $\sim 0.1\,\mathrm{dex}$), while it is slightly overestimated (by $\sim 0.2\,\mathrm{dex}$) for case B.
For both test cases the mean $1\,\sigma$-result overestimates the metallicity by slightly more than $0.1\,\mathrm{dex}$.
By the same amount the [Si/C] abundance is overestimated and underestimated in case A and B, respectively.
However, within the scatter all values are well reproduced.

The tightest constraints for the physical parameters are obtained by adopting the best-fit SEDs discussed in the previous Section.
We have found $\Delta_3 \simeq 0.0$ and $\Delta_3 \simeq -0.4$ for case A and B, respectively.
The results obtained when fixing these values and selecting the models with $1\,\sigma$ confidence are also listed in Table \ref{par_recover}. 
All parameters are reproduced within $2\,\sigma$, however, not necessarily to an improved precision compared to the previous selection, in particular noticeable for case A.

For the recovered SED with $(\Delta_1, \Delta_2, \Delta_3) = (0.0, 0.0, 0.0)$ and $(+0.4, -1.0, -0.4)$ for case A and B, respectively, the input values are well reproduced. 
Metallicity and [Si/C] are slightly overestimated (by $\lesssim 0.2\,\mathrm{dex}$) in case B, which can be explained by the slightly harder recovered SED compared to the original one.

Thus, the shape of the ionizing spectrum as well as the underlying physical quantities like ionization parameter (i.e.\ density), metallicity, and  relative abundance [Si/C] of the test cases are well recovered by our procedure.
We therefore are confident to obtain meaningful results also for observed data, provided that the SED of the ionizing radiation is, in principle, similar to the HM01 spectrum.

\section{Results from observations}\label{results}

The observed metal line systems presented in Sect.\ \ref{systems} are modeled in the same way as the test cases where for each system the model grid is based on the unmodified HM01 spectrum at the appropriate redshift.
The ionization parameter of the best-fit model for each test SED is determined by matching the observed \ion{Si}{iii}/\ion{Si}{iv} ratio.
In order to estimate the confidence of the fit we use in addition \ion{C}{iii}/\ion{C}{iv} for the systems at $z = 2.3799$ and $z = 2.8266$.
For the system at $z = 1.7529$ both \ion{Si}{ii}/\ion{Si}{iii} and \ion{Al}{ii}/\ion{Al}{iii} are considered (see below).
This means that at $z > 2$ we probe the same energy range as for the test cases.
At $z < 2$ the height of the 3\,Ryd Ly$\alpha$ peak, i.e.\ $\Delta_1$ is expected to have the strongest impact on the results since none of the ionization thresholds of the species taken into account are at $E > 4\,\mathrm{Ryd}$ (see Fig.\ \ref{uvb}).

\subsection{Constraints on the ionizing spectral energy distribution}

The confidence contours with respect to the parameterization of the SED projected into all possible planes are presented in Figs.\ \ref{projection_z>2} and \ref{projection_z<2}.
Obviously, the distributions of the best-fit models of the $z > 2$ systems (Fig.\ \ref{projection_z>2}) are similar to those found for the test cases while the low-redshift system appears to be exposed to a different spectrum (Fig.\ \ref{projection_z<2}).
This could be due to the different column density ratios adopted to evaluate the quality of the fit but a real evolution of the UV background is also possible \citep{fechner+2006a, agafonova+2007}.
Note, that the unmodified HM01 spectrum is not included into the confidence regions for none of the systems.

\subsubsection{$z > 2$}

\begin{figure*}
  \centering
   \resizebox{\hsize}{!}{\includegraphics[bb=0 330 590 565,clip=]{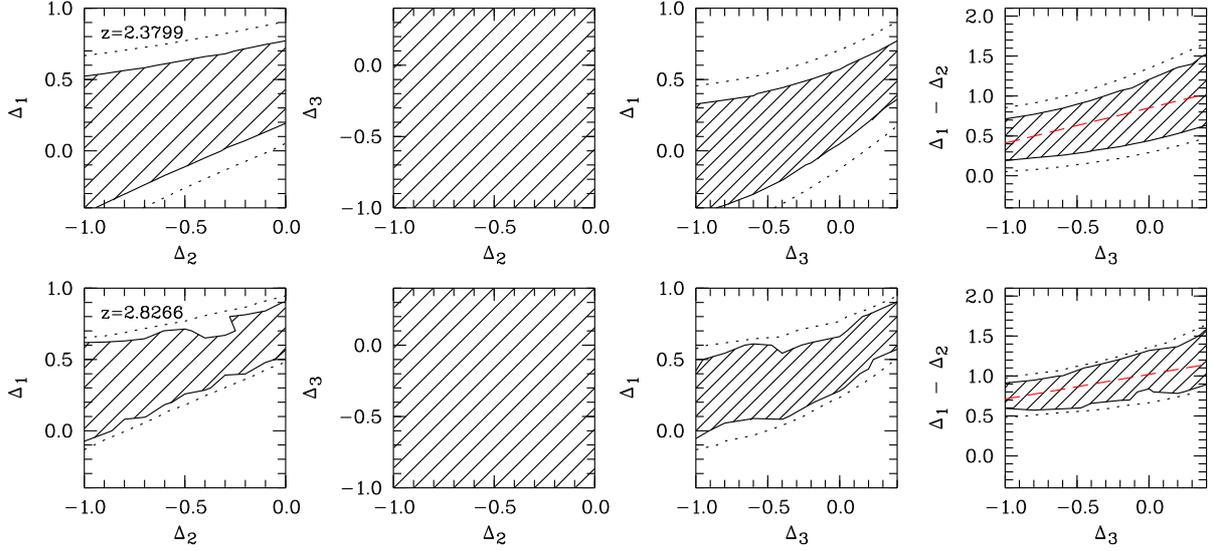}}
  \caption{Projected $1\,\sigma$ (marked areas) and $3\,\sigma$ (dotted lines) confidence contours of the SED parameters for the observed systems at $z = 2.3799$ (upper panels), and $z = 2.8266$ (lower panels).
The projections in each of the three possible planes are shown ($\Delta_1$ versus $\Delta_2$, $\Delta_3$ versus $\Delta_2$, and $\Delta_1$ versus $\Delta_3$; from left to right).
The right most panels display the confidence regions considering the enhancement of the break at 3\,Ryd, i.e.\ $\Delta_1 - \Delta_2$ versus the strength of the 4\,Ryd break $\Delta_3$, where the dashed lines indicate a fitted linear correlation (see text).
The contours of the $z=2.8266$ system are smoothed for a clearer presentation.
  }
  \label{projection_z>2}
\end{figure*}

\begin{figure}
  \centering
  \resizebox{\hsize}{!}{\includegraphics[bb=50 510 430 695,clip=]{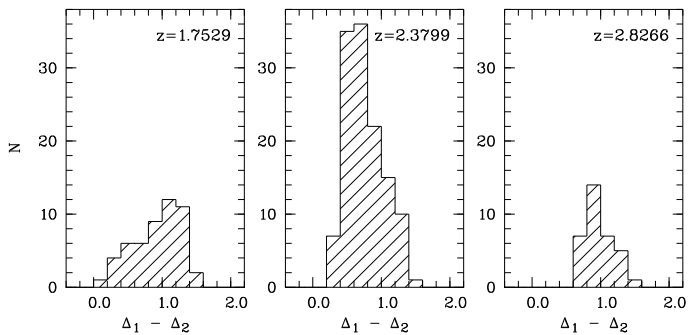}}
  \caption{Distribution of $\Delta_1 - \Delta_2$ for UV background spectra that fit the observed column density ratios up to $1\,\sigma$ confidence for the three observed systems at $z = 1.7529$ (left panel), $z = 2.3799$ (middle panel), and $z=2.8226$ (right panel).
  }
  \label{delta}
\end{figure}

In case of the $z > 2$ systems the SED parameters of the $1\,\sigma$ models are correlated in a similar way as those of the test cases.
For both systems more enhanced 3\,Ryd Ly$\alpha$ peaks, i.e.\ larger $\Delta_1$ are found for less depressed intensity in the 3-4\,Ryd range, i.e.\ larger $\Delta_2$.
In addition, $\Delta_1$ increases with increasing $\Delta_3$.
That means harder spectra favor more pronounced 3\,Ryd Ly$\alpha$ peaks.

The distribution of $\Delta_1 - \Delta_2$, which describes the intensity break at 3\,Ryd for the most confident models, are presented in the middle and right panel of Fig.\ \ref{delta}.
The mean values are found to be $\langle\Delta_1 - \Delta_2\rangle = 0.7\pm^{0.8}_{0.5}$ and $0.9\pm^{0.6}_{0.3}$ for the system at $z = 2.3799$ and $z=2.8266$, respectively.

Comparing the intensity break at 3\,Ryd $\Delta_1 - \Delta_2$ to the break at 4\,Ryd $\Delta_3$ in the right-most panels of Fig.\ \ref{projection_z>2} again yields a correlation between those parameters.
A linear fit leads to $\Delta_1 - \Delta_2 = (0.44\pm0.04)\cdot\Delta_3 + (0.85\pm0.02)$ for the system at $z = 2.3799$ and $\Delta_1 - \Delta_2 = (0.31\pm0.06)\cdot\Delta_3 + (1.02\pm0.03)$ for the system at $z = 2.8266$, respectively, indicated as dashed lines in Fig.\ \ref{projection_z>2}.
Analogously to the test cases, we compute rough estimates for the values $\Delta_3$ by inserting the mean $\Delta_1 - \Delta_2$ into the linear fits.
This results into $\Delta_3 \simeq -0.3$ for the system at $z=2.3799$ and $\Delta_3 \simeq -0.4$ for the system at $z=2.8266$.
The most confident models with this 4\,Ryd break are found at $(\Delta_1, \Delta_2, \Delta_3) = (+0.4, 0.0, -0.3)$ at $z = 2.3799$, where we have interpolated between the models with $\Delta_3 = -0.2$ and $-0.4$, and $(+0.3, -0.6, -0.4)$ at $z = 2.8266$, respectively.
Since the unmodified HM01 spectrum has been well reproduced by the test model and at least the correct tendencies of the modified UV background have been recovered, we are led to the conclusion that both systems at $z > 2$ are ionized by a spectrum different from the HM01 background.

\subsubsection{$z < 2$}

\begin{figure*}
  \centering
  \resizebox{\hsize}{!}{\includegraphics[bb=0 330 590 685,clip=]{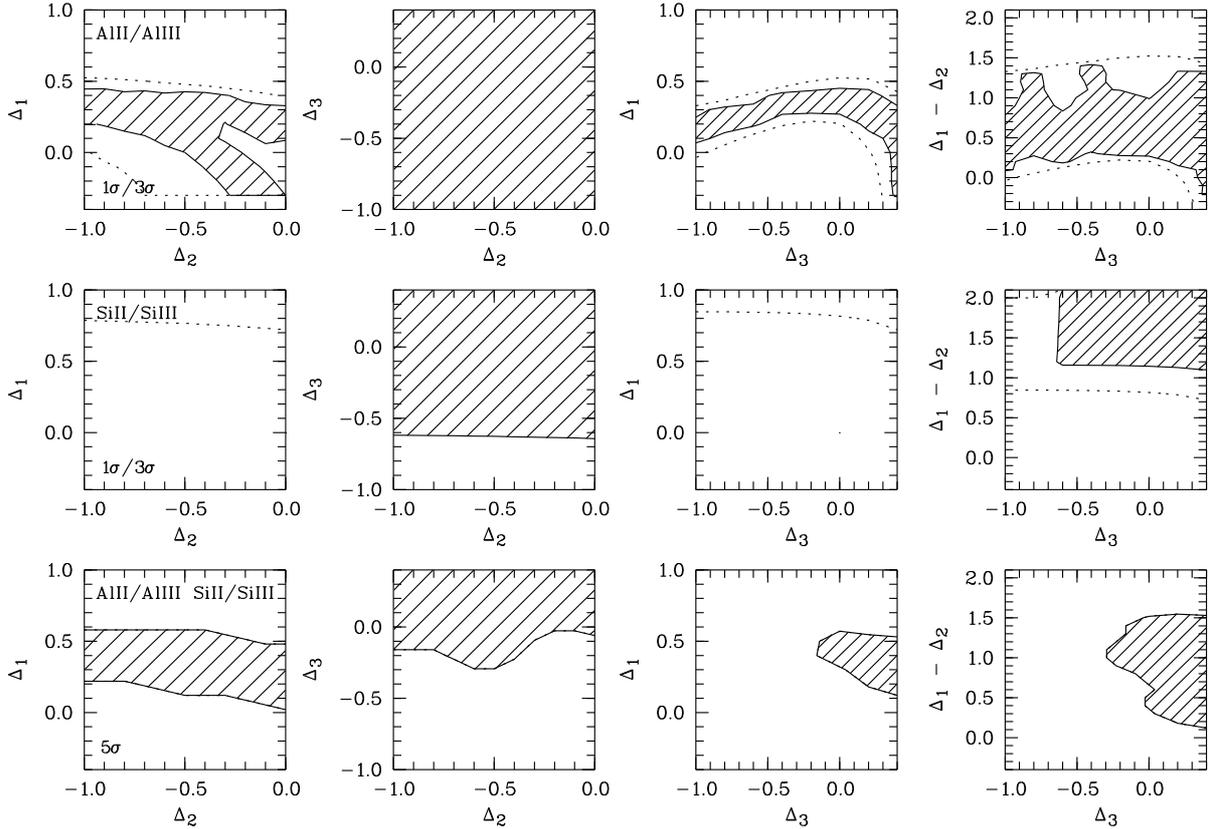}}
  \caption{Projected confidence contours of the SED parameters for the observed system at $z = 1.7529$ considering the \ion{Al}{ii}/\ion{Al}{iii} ratio (upper panels), the \ion{Si}{ii}/\ion{Si}{iii} ratio (middle panel), and the combination of both column density ratios (lower panel).
The projections in each of the three possible planes are shown ($\Delta_1$ versus $\Delta_2$, $\Delta_3$ versus $\Delta_2$, and $\Delta_1$ versus $\Delta_3$; from left to right).
The outer right panels display the confidence contours considering the enhancement of the break at 3\,Ryd, i.e.\ $\Delta_1 - \Delta_2$ versus the strength of the 4\,Ryd break $\Delta_3$.
The upper and middle panels show the $1\,\sigma$ (marked areas) and $3\,\sigma$ (dotted lines) ranges of confidence.
Since there are no SEDs that reproduce the \ion{Al}{ii}/\ion{Al}{iii} and \ion{Si}{ii}/\ion{Si}{iii} within $< 3\,\sigma$ simultaneously, the lower panels indicate the regions in parameter space where both column density ratios are matched within $5\,\sigma$ confidence
  }
  \label{projection_z<2}
\end{figure*}

The distribution of the parameters of the most confident models for the system at $z = 1.7526$ apparently differs from the $z > 2$  systems (Fig.\ \ref{projection_z<2}).
Since for this system two additional column density ratios are available, we first consider the $1\,\sigma$ and $3\,\sigma$ confidence regions for the \ion{Al}{ii}/\ion{Al}{iii} and \ion{Si}{ii}/\ion{Si}{iii} ratios separately (upper and middle panels of Fig.\ \ref{projection_z<2}, respectively).
Both ratios lead to rather tight constraints for the peak at 3\,Ryd, $\Delta_1$, since all considered column density ratios (\ion{Si}{iii}/\ion{Si}{iv}, \ion{Si}{ii}/\ion{Si}{iii}, and \ion{Al}{ii}/\ion{Al}{iii}) are sensitive to the SED at $E < 3\,\mathrm{Ryd}$.
While \ion{Al}{ii}/\ion{Al}{iii} requires $\Delta_1 \le +0.4$, the \ion{Si}{ii}/\ion{Si}{iii} ratio is only reproduced with $\sim 1\,\sigma$ confidence by models with $\Delta_1 = +1.0$.
Furthermore, \ion{Si}{ii}/\ion{Si}{iii} favors spectra with $\Delta_3 \gtrsim -0.4$ while \ion{Al}{ii}/\ion{Al}{iii} leaves $\Delta_3$ unconstrained.

Since neither the $1\,\sigma$ nor the $3\,\sigma$ ranges of \ion{Al}{ii}/\ion{Al}{iii} and \ion{Si}{ii}/\ion{Si}{iii} overlap, we consider a wider range of confident SEDs and mark in the lower panels of Fig.\ \ref{projection_z<2} those regions in parameter space that reproduce \ion{Al}{ii}/\ion{Al}{iii} and \ion{Si}{ii}/\ion{Si}{iii} simultaneously to a $5\,\sigma$ level of confidence.
Due to the higher uncertainties of the column densities of silicon, the number of confident models derived from the \ion{Si}{ii}/\ion{Si}{iii} ratio strongly increases when larger $\sigma$-ranges are considered.
Therefore, the joint $5\,\sigma$ confidence regions are mainly determined by the \ion{Al}{ii}/\ion{Al}{iii} constraints.
The 3\,Ryd Ly$\alpha$ peak is then found to be  $\langle\Delta_1\rangle = 0.4\pm^{0.1}_{0.3}$.
Best-fit values of the break at 3\,Ryd are in the range $\langle\Delta_1 - \Delta_2\rangle = 0.9\pm^{0.6}_{0.8}$.
Together with the well-constrained result for $\Delta_1$ we estimate $\Delta_2 \sim -0.5$.
Furthermore, hard spectra with $\Delta_3 > -0.2$ are favored.
The actual best-fit model is yielded at $(\Delta_1, \Delta_2, \Delta_3) = (+0.4, -0.5, +0.4)$ suggesting that the system at $z = 1.7529$ favors a hard ionizing radiation.

\subsection{Physical parameters}

\begin{figure}
  \centering
   \resizebox{\hsize}{!}{\includegraphics[bb=48 27 285 270,clip=,angle=-90]{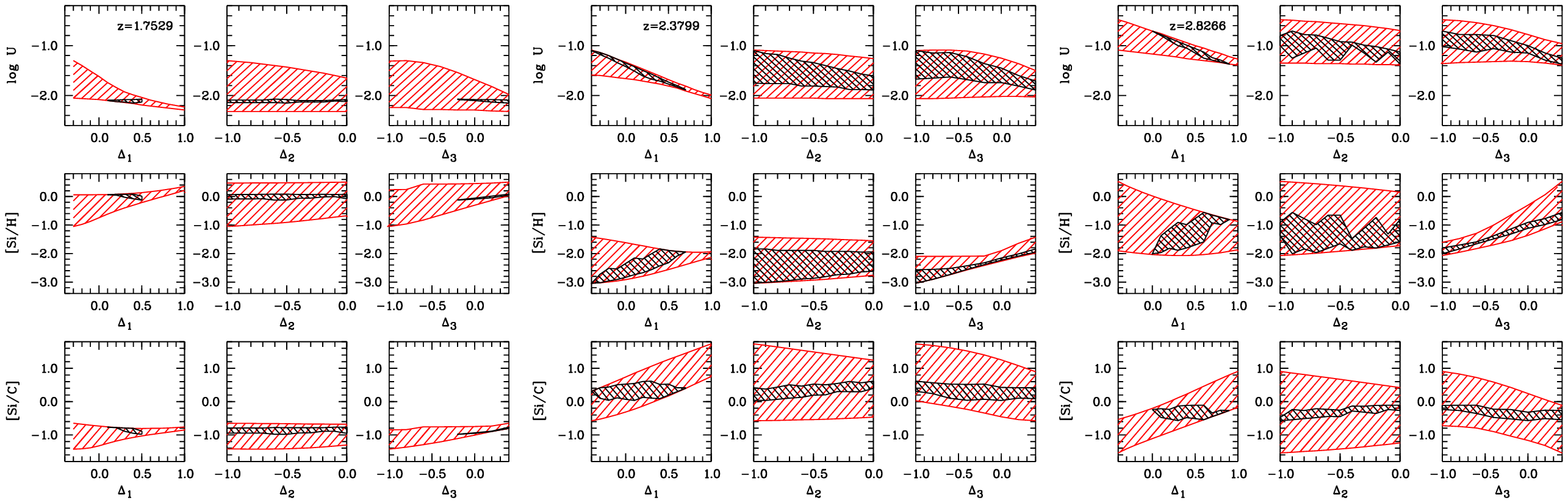}}
  \caption{Dependence of physical parameters on the adopted ionizing spectrum for the observed low-redshift system at $z = 1.7529$.
The wide and narrow marked areas in the upper, middle, and lower panels show the ionization parameter $\log U$, metallicity [Si/H], and relative abundance [Si/C], respectively, versus the parameters of the ionizing SED ($\Delta_1$, $\Delta_2$, $\Delta_3$; from the left to the right) for all best-fit models and those consistent with the observed column density ratios of \ion{Al}{ii}/\ion{Al}{iii} and \ion{Si}{ii}/\ion{Si}{iii} on a $5\,\sigma$ level, respectively.
  }
  \label{model_par1}
\end{figure}

\begin{figure*}
  \centering
   \resizebox{\hsize}{!}{\includegraphics[bb=48 275 285 765,clip=,angle=-90]{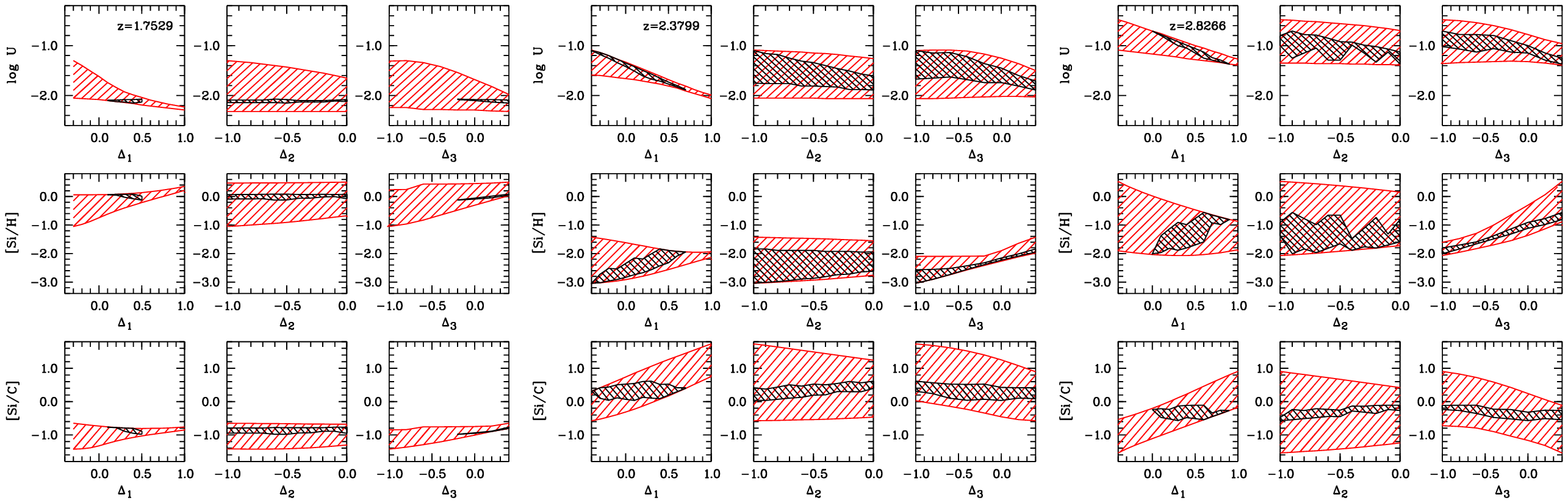}}
  \caption{Dependence of physical parameters on the adopted ionizing spectrum for the observed systems at $z > 2$.
The marked areas in the upper, middle, and lower panels show the ionization parameter $\log U$, metallicity [Si/H], and relative abundance [Si/C], respectively, versus the parameters of the ionizing SED ($\Delta_1$, $\Delta_2$, $\Delta_3$; from the left to the right) for the best-fit models of the systems at $z = 2.3799$ (left) and $z = 2.8266$ (right).
The narrower areas indicate the parameters of the models consistent with the observed \ion{C}{iii}/\ion{C}{iv} column density ratios on a $1\,\sigma$ level.
  }
  \label{model_par2}
\end{figure*}

\begin{table}
  \caption[]{Results for the three studied metal line systems.
}
  \label{par_sys}
  $$
  \begin{array}{c l l l l l}
    \hline\hline
    \noalign{\smallskip}
z_{\mathrm{sys}} & \mathrm{considered~models} &\log n_{\element{H}}& \log U  &\mathrm{[Si/H]}&\mathrm{[Si/C]}\\
    \noalign{\smallskip}
    \hline
    \noalign{\smallskip}
1.7529  &\mathrm{all~models} &-2.46\pm^{0.51}_{0.83}&-1.99\pm^{0.69}_{0.30}&-0.16\pm^{0.52}_{0.89}&-0.99\pm^{0.34}_{0.44}\\
 &5\,\sigma &-2.34\pm^{0.05}_{0.06}&-2.11\pm^{0.04}_{0.04}&+0.01\pm^{0.08}_{0.14}&-0.86\pm^{0.09}_{0.13}\\
 & \mathrm{recovered~SED}^{\,\mathrm{a}} & -2.30 & -2.15 & +0.08 & -0.81 \\
 & \mathrm{HM01} & -2.61 & -1.92 & -0.22 & -0.99 \\
    \noalign{\smallskip}
2.3799 &\mathrm{all~models} & -2.83\pm^{0.65}_{0.69}&-1.63\pm^{0.54}_{0.44}&-2.28\pm^{0.86}_{0.76}&+0.46\pm^{1.28}_{1.04}\\
 &1\,\sigma &-2.97\pm^{0.52}_{0.54}&-1.52\pm^{0.42}_{0.37}&-2.41\pm^{0.57}_{0.63}&+0.32\pm^{0.30}_{0.28}\\
 & 1\,\sigma\,(\Delta_3=-0.3) &-3.04\pm^{0.28}_{0.23}&-1.47\pm^{0.18}_{0.21}&-2.44\pm^{0.05}_{0.03}&+0.27\pm^{0.22}_{0.22}\\
 & \mathrm{recovered~SED}^{\,\mathrm{b}}&-2.77 & -1.67 & -2.39 & +0.46\\
 &\mathrm{HM01} & -3.02 & -1.50 & -2.12 & -0.11 \\
    \noalign{\smallskip}
2.8266  &\mathrm{all~models} &-3.43\pm^{0.61}_{0.70}&-1.01\pm^{0.53}_{0.38}&-1.20\pm^{1.72}_{0.87}&-0.39\pm^{1.29}_{1.16}\\
 &1\,\sigma &-3.39\pm^{0.50}_{0.44}&-1.04\pm^{0.33}_{0.33}&-1.31\pm^{0.75}_{0.70}&-0.33\pm^{0.22}_{0.24}\\
 & 1\,\sigma\,(\Delta_3=-0.4) &-3.54\pm^{0.19}_{0.18}&-0.93\pm^{0.13}_{0.13}&-1.45\pm^{0.07}_{0.04}&-0.35\pm^{0.12}_{0.13}\\
 & \mathrm{recovered~SED}^{\,\mathrm{c}}&-3.54 & -0.93 & -1.46 & -0.34\\
 & \mathrm{HM01} & -3.52 & -1.00 & -0.73 & -0.79 \\
    \noalign{\smallskip}
    \hline
  \end{array}
  $$
  \begin{list}{}{}
    \item[$^{\mathrm{a}}$] recovered SED has $(\Delta_1, \Delta_2, \Delta_3) = (+0.4, -0.5, +0.4)$
    \item[$^{\mathrm{b}}$] recovered SED has $(\Delta_1, \Delta_2, \Delta_3) = (+0.4, 0.0, -0.3)$
    \item[$^{\mathrm{c}}$] recovered SED has $(\Delta_1, \Delta_2, \Delta_3) = (+0.3, -0.6, -0.4)$
  \end{list}
\end{table}

The estimated values for the ionization parameter $\log U$, the metallicity [Si/H], and the relative abundance [Si/C] versus the SED parameters are shown in Figs.\ \ref{model_par1} and \ref{model_par2} and summarized in Table \ref{par_sys}.
The marked narrow areas in Fig.\ \ref{model_par1} indicate the range for those models of the system at $z=1.7529$ that fit various measured column density ratios simultaneously within $5\,\sigma$ confidence, while the narrow areas in Fig.\ \ref{model_par2} mark the $1\,\sigma$ confidence regions for the systems at $z > 2$ based on the \ion{C}{iii}/\ion{C}{iv} ratio.
For the system at $z = 1.7529$ (Fig.\ \ref{model_par1}) these models yield physical parameters in an extremely narrow range.
On average we find $\langle\log U\rangle = -2.11 \pm 0.04$ corresponding to $\langle\log n_{\element{H}}\rangle = -2.34\pm^{0.05}_{0.06}$, $\langle\mathrm{[Si/H]}\rangle = +0.01 \pm ^{0.08}_{0.14}$, and $\langle\mathrm{[Si/C]}\rangle = -0.86\pm^{0.09}_{0.13}$ where the error bars denote the range of estimated values.
Adopting the HM01 background leads to $\log U \simeq -1.92$ and $\mathrm{[Si/H]} \simeq -0.22$, roughly $0.2\,\mathrm{dex}$ higher and lower than the mean of the confident models reflecting that those spectra are considerably harder than HM01.

The corresponding plots for the $z = 2.3799$ and $z=2.8266$ systems are given in the left and right panels of Fig.\ \ref{model_par2}, respectively.
Due to the higher uncertainty of the \ion{C}{iii} column density a wider range of consistent models and parameter values is found.
For the $z = 2.3799$ system we derive $\langle\log U\rangle = -1.52 \pm^{0.42}_{0.37}$ for the ionization parameter and a metallicity of $\langle\mathrm{[Si/H]}\rangle = -2.41 \pm ^{0.57}_{0.63}$ with $1\,\sigma$ confidence. 
The results derived for a HM01 spectrum ($\log U \simeq -1.50$ and $\mathrm{[Si/H]} \simeq -2.12$) are within this range although the HM01 model is not within the $1\,\sigma$ confidence range.
The system at $z = 2.8266$ yields $\langle\log U\rangle = -1.04 \pm 0.33$ and $\langle\mathrm{[Si/H]}\rangle = -1.31 \pm ^{0.75}_{0.70}$ also consistent with the HM01 result of $\log U \simeq -1.00$ and $\mathrm{[Si/H]} \simeq -0.73$ even though the HM01 spectrum significantly underestimates the observed \ion{C}{iii}/\ion{C}{iv} ratio.
Note that already the $1\,\sigma$ results for the metallicity and hydrogen density of the $z > 2$ systems are spread over more than one order of magnitude strongly depending on the hardness of the ionizing radiation and the strength of the Ly$\alpha$ peak at 3\,Ryd.

Following the analysis procedure applied to the test cases, we fix $\Delta_3$ to the recovered value, which is $-0.3$ and $-0.4$, respectively, and select the models fitting at least with  $1\,\sigma$ confidence.
The ranges of the physical parameters then strongly reduce (see Table \ref{par_sys}).
In particular, the metallicity is now tightly constrained with a spread of $\Delta \mathrm{[Si/H]} \sim 0.1$.
The results for the recovered SEDs are also listed in Table \ref{par_sys}.
Compared to the HM01 result, we find only slightly different ionization parameters.
But for both $z > 2$ systems the metallicity derived from the recovered SED is lower by $\sim 0.3$ to $0.7\,\mathrm{dex}$ due to the softer spectrum.
Furthermore, the enhancement of the 3\,Ryd Ly$\alpha$ peak leads to a higher [Si/C] abundance by roughly $0.5\,\mathrm{dex}$.

\section{Discussion}\label{discussion}

\begin{figure}
  \centering
   \resizebox{\hsize}{!}{\includegraphics[bb=30 225 310 740,clip=]{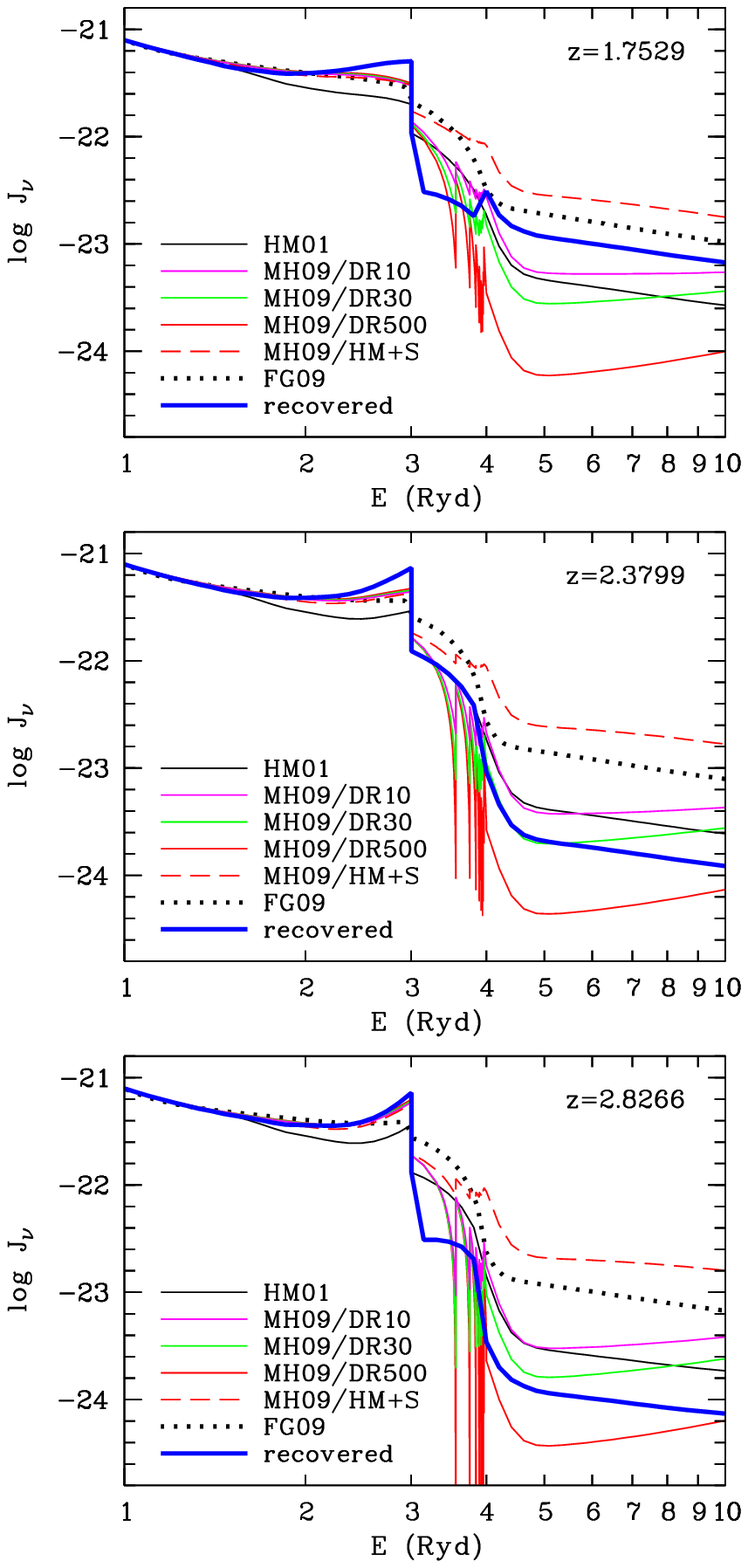}}
  \caption{Recovered SEDs in comparison to UV background spectra from the literature for the metal line systems at $z = 1.7529$ (upper panel), $z = 2.3799$ (middle panel), and $z = 2.8266$ (lower panel).
    Thick blue lines display our fiducial spectra described by $(\Delta_1, \Delta_2, \Delta_3) = (+0.4, -0.5, +0.4)$ at $z = 1.7529$, $(+0.4, 0.0, -0.3)$ at $z = 2.3799$, and $(+0.3, -0.6, -0.4)$ at $z = 2.8266$, respectively, compared to the unmodified \citet{haardt+2001} background (thin black line; HM01), the sawtooth spectrum of \citet{madau+2009} in the \citet{haardt+1996} uniform background
limit (red dashed line; MH09/HM+S) and with delayed helium reionization boosting the \ion{He}{ii}/\ion{H}{i} by factors of 10, 30, and 500 compared to the MH09/HM+S case (thin solid lines; MH09/DR10, MH09/DR30, MH09/DR500), respectively, and the UV background spectrum of \citet[][dotted line; FG09]{faucher+2009} at the appropriate redshifts.
All spectra are normalized to $\log J = -21.1$ at 1\,Ryd.
  }
  \label{uvb_comp}
\end{figure}

As shown in the previous Section, evaluating best-fit photoionization models of three observed metal line systems suggest that the ionizing radiation differs from the usually adopted \citet{haardt+2001} background spectrum.
In two systems at $z > 2$ a softer spectrum has been recovered where the \ion{He}{ii} re-emission peak at 3\,Ryd is slightly enhanced and the intensity level in the range 3-4\,Ryd is reduced.
Measuring the deviation from the HM01 spectrum in dex, our fiducial SEDs can be parameterized with $(\Delta_1, \Delta_2, \Delta_3) = (+0.4, 0.0, -0.3)$ at $z = 2.3799$ and $(+0.3, -0.6, -0.4)$ at $z = 2.8266$, respectively. 
For one system at $z = 1.7529$ a hard spectrum even more different from HM01 is suggested yielding $(+0.4, -0.5, +0.4)$ as fiducial SED.

\subsection{The spectrum of the UV background}

Our fiducial spectra are shown in Fig.\ \ref{uvb_comp} as thick solid lines.
Keep in mind, that the uncertainties of these SEDs are high and hardly to quantify since the parameters are mutually correlated.
Nevertheless, in the following we compare our best-fit recovered SEDs to various UV background spectra from the literature, which are the classical \citet{haardt+2001} spectrum, various upgradings based on \citet{madau+2009}, and the independently computed background from \citet[][see Fig.\ \ref{uvb_comp}]{faucher+2009}.

The recent model of \citet[][FG09]{faucher+2009} at the appropriate redshift is indicated as dotted line in each panel of Fig.\ \ref{uvb_comp}.
At each redshift the FG09 background is significantly flatter than our fiducial SEDs and even slightly flatter than the pure HM01 background (thin solid black lines).
The FG09 spectra show only very weak recombination emission at 3\,Ryd.
According to the authors this is due to the combined effects of the saturation of re-emission in optically thick systems, leakage of the re-emitted photons at the ionizing edge, and the frequency dependence of the photoionization cross section.
In addition to the weak 3\,Ryd Ly$\alpha$ peak the FG09 spectra show a high intensity in the 3-4\,Ryd range, leading to a weak break at 3\,Ryd, which is even less pronounced than in the HM01 spectrum (thin black line in Fig.\ \ref{uvb_comp}).
Since this feature is robustly constrained by our models to be larger than in the HM01 case, we conclude that the \ion{He}{ii} Lyman continuum absorption and recombination re-emission may be underestimated in the FG09 models.

Furthermore, the break at 4\,Ryd is weaker in the FG09 spectra than in the HM01 background.
Consequently, the FG09 spectra provide a harder radiation field.
As demonstrated by \citet{faucher+2009} the depth of the 4\,Ryd break mainly depends on the slope $\beta$ of the \ion{H}{i} column density distribution and the spectral index $\alpha_{\mathrm{QSO}}$ of the contributing QSOs.
The steeper the column density distribution or the softer the QSO radiation, the stronger the break at the \ion{He}{ii} edge.
In both cases the number of \ion{He}{ii} Lyman limit systems is increased leading to more absorption just above 4\,Ryd.
Thus, comparison to our fiducial spectra at $z > 2$ suggest that either the \ion{H}{i} column density distribution slope is steeper than the adopted $\beta = 1.4$ of \citet{faucher+2009} or the contribution of QSOs to the UV background at these redshifts is less than assumed.

The recent finding of a flat, i.e.\ nearly constant, hydrogen photoionization rate at  $z > 2$ by \citet{faucher+2008c} and \citet{dallaglio+2009} suggests that stars contribute significantly to the ionizing radiation at $z \gtrsim 3$.
If galaxies dominate the UV background, a softer SED is expected.
The studied systems may probe the transition from a QSO- to a galaxy-dominated background in the redshift range $2 \lesssim z \lesssim 3$.
Alternatively, the absorbers may be affected by an additional contribution from local soft radiation sources \citep{miralda2005, schaye2006}.

Recently, \citet{madau+2009} have shown that resonant line absorption in the Lyman series of \ion{He}{ii} leads to a sawtooth modulation of the UV background spectrum between 3 and 4\,Ryd.
The resulting energy distributions at the appropriate redshifts are also shown in Fig.\ \ref{uvb_comp} as dashed red lines (MH09/HM+S).
In this model only QSOs are considered as sources of radiation and photoionization equilibrium with a uniform radiation field is assumed.
Since the soft radiation contributed by galaxies is neglected, the sawtooth SED is much harder than the HM01 background which takes into account galaxy spectra as well.

A very recent upgrade of the HM spectrum \citep[][HM11]{haardt+2011} includes the sawtooth modulation as well as the radiation of galaxies \citep[for more details we refer to][]{haardt+2011}.
A rough comparison of the new model to the HM01 background at $z \sim 3$ shows that besides the sawtooth modulation the HM11 spectrum declines slightly more steeply between 1 and 3\,Ryd and has a weaker break at 4\,Ryd.
At higher energies, the intensity of the HM11 spectrum drops significantly, in particular at $E \gtrsim 45\,\mathrm{Ryd}$, while the HMN01 background is still flat in this energy regime.

The colored thin solid lines in Fig.\ \ref{uvb_comp} represent a sawtooth model according to \citet{madau+2009} which additionally assumes late \ion{He}{ii} reionization by artificially boosting the \ion{He}{ii}/\ion{H}{i} ratio by a factor of 10, 30, and 500 (MH09/DR10, MH09/DR30, MH09/DR500), respectively.
In these cases, the spectrum is suppressed above 4\,Ryd and the sawtooth modulation results in a net depression of the intensity level in the range 3-4\,Ryd.
Both effects increase with the amount of existing \ion{He}{ii}, i.e.\ with the \ion{He}{ii}/\ion{H}{i} ratio.
Furthermore, the re-emission peak at 3\,Ryd is enhanced.
Thus, at $z > 2$ the MH09/DR spectra show similar overall features as our recovered SEDs.
In particular, for the system at $z = 2.3799$ the recovered spectrum near 4\,Ryd follows closely the MH09/DR30 spectrum (middle panel of Fig.\ \ref{uvb_comp}) and for the system at $z = 2.8266$ the 3\,Ryd Ly$\alpha$ peak of the recovered SED and all the MH09/DR spectra are nearly identical (lower panel of Fig.\ \ref{uvb_comp}).

\begin{figure}
  \centering
  \resizebox{\hsize}{!}{\includegraphics[bb=40 335 360 755,clip=]{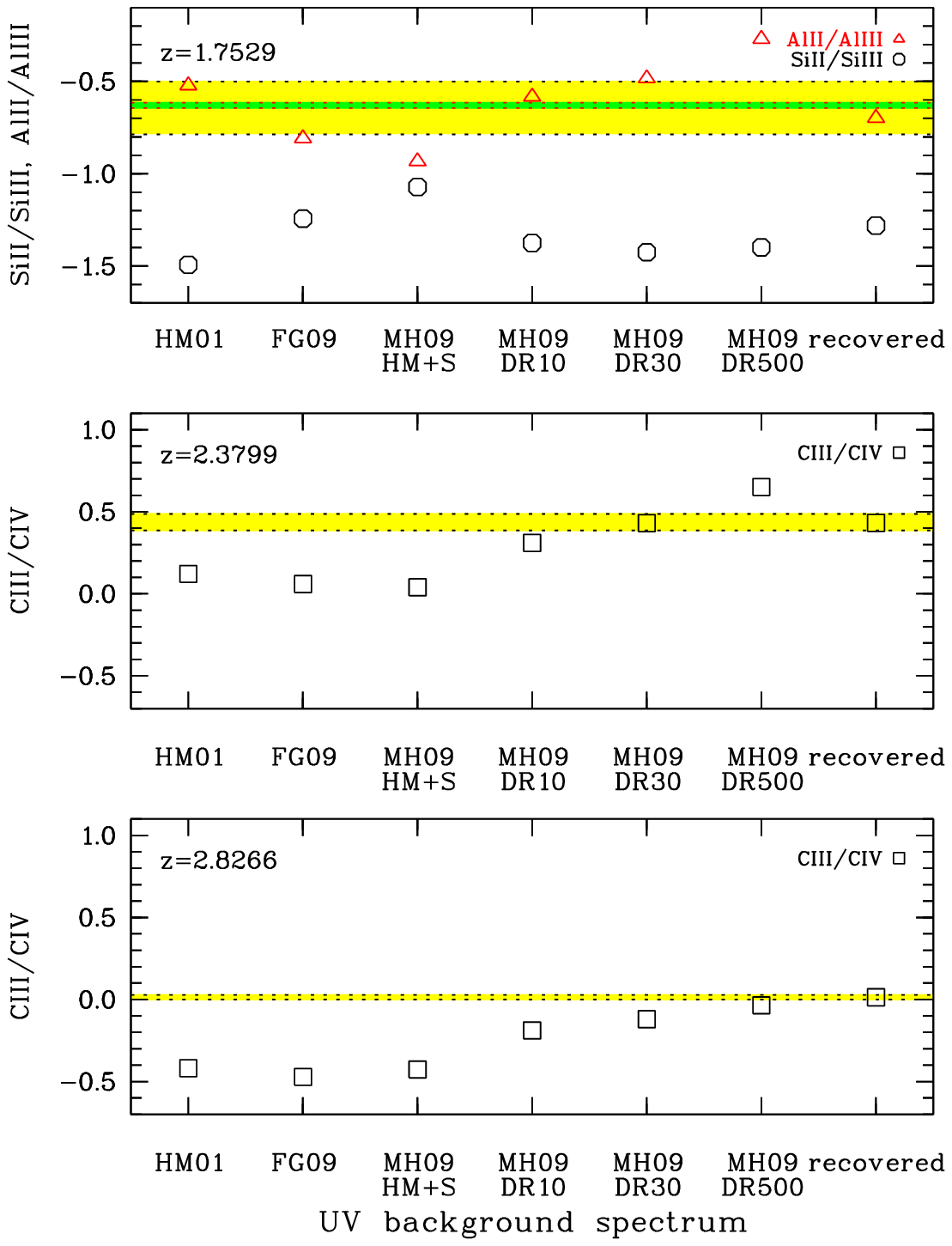}}
  \caption{Comparison of predicted column density ratios to the observations for various UV background models from the literature.
The adopted ionizing spectra are from \citet[][HM01]{haardt+2001}, \citet[][FG09]{faucher+2009}, \citet{madau+2009} assuming equilibrium conditions (MH09/HM+S) or delayed helium reionization with \ion{He}{ii}/\ion{H}{i} boosted by a factor 10, 30, and 500 (MH09/DR10, MH09/DR30, MH09/DR500), respectively, and our recovered SEDs (see text).
Shaded areas mark the observed ratio of \ion{Si}{ii}/\ion{Si}{iii} (upper panel, wider range) and \ion{Al}{ii}/\ion{Al}{iii} (upper panel, narrow range) for the $z = 1.7529$ system and \ion{C}{iii}/\ion{C}{iv} for the systems at $z = 2.3799$ (middle panel) and $z = 2.8266$ (lower panel).
  }
  \label{uvb_comp2}
\end{figure}

\begin{table}
  \caption[]{Comparison of UV background spectra from the literature.
}
  \label{par_uvb}
  $$
  \begin{array}{c l l l l l l r}
    \hline\hline
    \noalign{\smallskip}
z_{\mathrm{sys}} & \mathrm{ionizing~spectrum}^{\,\mathrm{d}} &\log n_{\element{H}}& \log U  &\mathrm{[Si/H]}&\mathrm{[Si/C]} & \log\eta & \sigma\\
    \noalign{\smallskip}
    \hline
    \noalign{\smallskip}
1.7529 & \mathrm{recovered~SED}^{\,\mathrm{a}} & -2.30 & -2.15 & +0.08 & -0.81 & 1.61 & 4.7 \\
 & \mathrm{HM01} & -2.61 & -1.92 & -0.22 & -0.99 & 1.95 & 6.8 \\
 & \mathrm{FG09} & -2.27 & -2.17 & +0.15 & -0.76 & 1.50 & 8.5 \\
 & \mathrm{MH09/HM+S} & -2.16 & -2.27 & +0.31 & -0.61 & 1.29 & 12.4 \\
 & \mathrm{MH09/DR10} & -2.37 & -2.08 & +0.01 & -0.84 & 1.78 & 4.2 \\
 & \mathrm{MH09/DR30} & -2.44 & -2.01 & -0.09 & -0.91 & 1.96 & 7.9 \\
 & \mathrm{MH09/DR500} & -2.77 & -1.68 & -0.46 & -1.15 & 2.51 & 15.5 \\
    \noalign{\smallskip}
2.3799 & \mathrm{recovered~SED}^{\,\mathrm{b}}&-2.77 & -1.67 & -2.39 & +0.46 & 2.37 & 0.1 \\
 & \mathrm{HM01} & -3.02 & -1.50 & -2.12 & -0.11 & 2.20 & 6.3 \\
 & \mathrm{FG09} & -2.71 & -1.72 & -1.81 & -0.17 & 1.81 & 7.5 \\
 & \mathrm{MH09/HM+S} & -2.56 & -1.87 & -1.53 & -0.24 & 1.50 & 7.9 \\
 & \mathrm{MH09/DR10} & -2.88 & -1.57 & -1.81 & -0.22 & 2.20 & 2.5 \\
 & \mathrm{MH09/DR30} & -2.94 & -1.51 & -1.96 & -0.10 & 2.47 & 0.1 \\
 & \mathrm{MH09/DR500} & -2.96 & -1.49 & -2.51 & +0.41 & 2.98 & 4.3 \\
    \noalign{\smallskip}
2.8266 & \mathrm{recovered~SED}^{\,\mathrm{c}}&-3.54 & -0.93 & -1.46 & -0.34 & 2.79 & 0.1 \\
 & \mathrm{HM01} & -3.52 & -1.00 & -0.73 & -0.79 & 2.34 & 30.9 \\
 & \mathrm{FG09} & -3.17 & -1.26 & -0.50 & +0.83 & 1.89 & 34.7 \\
 & \mathrm{MH09/HM+S} & -2.94 & -1.49 & -0.09 & -0.85 & 1.53 & 31.5 \\
 & \mathrm{MH09/DR10} & -3.32 & -1.13 & -0.24 & -1.05 & 2.27 & 14.5 \\ 
 & \mathrm{MH09/DR30} & -3.43 & -1.02 & -0.33 & -1.05 & 2.52 & 9.6 \\ 
 & \mathrm{MH09/DR500} & -3.57 & -0.87 & -1.02 & -0.72 & 3.13 & 3.6 \\ 
    \noalign{\smallskip}
    \hline
  \end{array}
  $$
  \begin{list}{}{}
    \item[$^{\mathrm{a}}$] recovered SED has $(\Delta_1, \Delta_2, \Delta_3) = (+0.4, -0.5, +0.4)$
    \item[$^{\mathrm{b}}$] recovered SED has $(\Delta_1, \Delta_2, \Delta_3) = (+0.4, 0.0, -0.3)$
    \item[$^{\mathrm{c}}$] recovered SED has $(\Delta_1, \Delta_2, \Delta_3) = (+0.3, -0.6, -0.4)$
    \item[$^{\mathrm{d}}$] HM01: \citet{haardt+2001}, FG09: \citet{faucher+2009}, MH09/HM+S: \citet{madau+2009} photoionization equilibrium model, MH09/DR10, MH09/DR30, MH09/DR500: \citet{madau+2009} delayed helium reionization model where the \ion{He}{ii}/\ion{H}{i} ratio is boosted by a factor 10, 30, and 500, respectively
  \end{list}
\end{table}

For each UV background spectrum from the literature we compute with CLOUDY a best-fit photoionization model for each observed system.
The resulting physical parameters are summarized in Table \ref{par_uvb}.
In addition to the hydrogen density $\log n_{\element{H}}$, the ionization parameter $\log U$, the metallicity [Si/H], and the relative abundance [Si/C], the \ion{He}{ii}/\ion{H}{i} ratio $\log\eta$ is listed.
The \ion{He}{ii}/\ion{H}{i} ratio measures the hardness of the ionizing radiation which is mainly determined by the intensity level above 4\,Ryd.
Therefore, the strongly absorbed MH09/DR500 background yields the highest $\eta$-values indicating a soft radiation field while the relatively flat MH09/HM+S and FG09 spectra lead to lower values corresponding to hard spectra.
In order to evaluate the quality of the models we compare the predicted column density ratios to the observed values in Fig.\ \ref{uvb_comp2}.
As above the \ion{C}{iii}/\ion{C}{iv} ratio is adopted at $z > 2$ (lower panels) while \ion{Si}{ii}/\ion{Si}{iii} and \ion{Al}{ii}/\ion{Al}{iii} are used at $z < 2$ (upper panel).

In case of the low-redshift system the quality of the fit is dominated by the \ion{Al}{ii}/\ion{Al}{iii} ratio (indicated as triangles) which is tightly constrained by the observations (narrow marked range).
Besides our fiducial model, the MH09/DR10 spectrum fits the observed column density ratios with $4.2\,\sigma$ confidence.
Though our recovered SED suggests a hard ionizing radiation, the models based on the harder backgrounds FG09 and MH09/HM+S are less accurate.
However, the selected $5\,\sigma$-confidence models cover the range of hardness $\log\eta = 1.70\pm^{0.20}_{0.11}$ corresponding to $\eta \approx 50\pm^{30}_{10}$.
The FG09 and MH09/HM+S spectra are even harder yielding $\eta \sim 30$ and $20$, respectively, while the $\eta$-value of the MH09/DR10 background ($\sim 60$) matches the range of the recovered spectra.
Comparing to direct measurements of the \ion{He}{ii}/\ion{H}{i} ratio at $z \gtrsim 2$ \citep{zheng+2004, fechner+2006b, fechner+2007a, shull+2010} shows that the recovered $\eta$-values are at the hard end of the expected range.

A hardening of the UV background at $z<2$ has been claimed already by other studies \citep{fechner+2006a, agafonova+2007}.
However, for the system at $z = 1.7529$ the results should be interpreted very cautiously.
Absorption features of various species are detected in this system (see Fig.\ \ref{profiles_z1.7529}). 
In particular, the system exhibits strong lines of \ion{C}{iv} and \ion{C}{ii} simultaneously.
Due to their different ionization potentials ($1.79\,\mathrm{Ryd}$ for \ion{C}{ii} and $4.74\,\mathrm{Ryd}$ for \ion{C}{iv}) both species probably do not live in the same gas phase.
While the \ion{C}{ii} absorption is thought to arise from the same phase as \ion{Si}{ii}, \ion{Si}{iii}, \ion{Al}{ii}, and \ion{Al}{iii}, \ion{C}{iv} is expected to trace a more highly-ionized gas phase.
Possibly, part of the \ion{Si}{iv} absorption may arise from the \ion{C}{iv} phase as well.
If so, the measured \ion{Si}{iii}/\ion{Si}{iv} ratio used to constrain the models, would actually be a lower limit and, therefore, our results would be biased.
Thus, an independent confirmation for a $z < 2$ metal line system using \ion{C}{iii}/\ion{C}{iv} is desirable.
But due to the short restwavelength of \ion{C}{iii} ($977.0201\,\mathrm{\AA}$) this requires high-quality near-UV observations.

At $z > 2$ our fiducial SEDs optimally reproduce the observed \ion{C}{iii}/\ion{C}{iv} ratio since they are constructed to do so.
Furthermore, the delayed reionization sawtooth models lead to surprisingly good fits for both systems.
The column densities of the system at $z = 2.3799$ are optimally modeled by the MH09/DR30 spectrum.
The other delayed reionization models lead to acceptable fits as well reproducing the \ion{C}{iii}/\ion{C}{iv} ratio within $2.5\,\sigma$ (MH09/DR10) and $4.7\,\sigma$ (MH09/DR500), respectively.
Note, that the physical parameters derived when adopting the MH09/DR30 spectrum differ significantly from those estimated with our recovered SED by upto $0.5\,\mathrm{dex}$ (see Table \ref{par_uvb}).
However, the values are consistent with the $1\,\sigma$ range listed in Table \ref{par_sys}.
The system at $z = 2.8266$ is fitted best by the MH09/DR500 spectrum ($3.6\,\sigma$ confidence).
Though the amount of \ion{He}{ii} required to generate such a soft radiation field seems unrealistically high, we conclude that the ionizing radiation should indeed be rather soft.

At least for these two individual systems our findings are contrary to \citet{vasiliev+2010}.
They have compared the observed \ion{C}{iii}/\ion{C}{iv} ratio of 10 absorbers to (non-)equilibrium photoionization models for several UV background spectra from \citet{madau+2009}. 
The authors come to the conclusion, that the delayed reionization models (MH09/DR) and thus a large fraction of \ion{He}{ii} is disfavoured by their data.

\subsection{Physical parameters}

The physical parameters of the analyzed metal line systems are also given in Table \ref{par_uvb}.
Though the various adopted ionizing spectra lead to different results for ionization parameter, i.e.\ hydrogen density, metallicity [Si/H], and [Si/C] abundance, the derived values indicate distinct results for the three systems.

The system at $z = 1.7529$ shows roughly solar [Si/H] with silicon depleted with respect to carbon with $\mathrm{[Si/C]} \sim -0.8$ for all adopted spectra.
This means, the system is apparently metal-rich with $\mathrm{[C/H]} \sim +0.8$.
The system at $z = 2.3799$ has sub-solar metallicity, $\mathrm{[Si/H]} \sim \mathrm{[C/H]} \sim -2.0$, with roughly solar [Si/C].
The soft recovered SED and the even softer MH09/DR500 spectrum lead to enhanced silicon abundance of $\mathrm{[Si/C]} \sim +0.4$ as would be expected in the IGM \citep{aguirre+2008}.
The system at $z = 2.8266$ appears to be metal-rich as well.
For the adopted spectra we find about solar metallicity, $\mathrm{[C/H]} \sim -0.1$, and silicon being depleted with respect to carbon by $\mathrm{[Si/C]} \sim -0.6$.
Note, that the hard UV background of FG09 suggests enhanced silicon for this system (see Table \ref{par_uvb}).

Two of the analyzed systems have been investigated in detail by \citet[][$z = 2.3799$ towards HS~1700+6416]{agafonova+2007} and \citet[][$z = 1.7925$ towards HE~1347-2457]{levshakov+2009}, respectively.
For both systems the measured column densities are generally in good agreement.
In case of the $z = 2.3799$ system, \citet{agafonova+2007} also estimate similar abundances ($\mathrm{[C/H]} = -1.94$ and $\mathrm{[Si/C]} = +0,33$) even though they applied a slightly different UV background spectrum, 

However, for the system at $z = 1.7529$ \citet{levshakov+2009} estimate a significantly higher \ion{H}{i} column density ($\log N(\ion{H}{i}) \simeq 16.0$ in comparison to our $\sim 15.5$).
Since the Ly$\alpha$ feature is saturated and higher order Lyman series lines are not observed, the \ion{H}{i} column density cannot be estimated unambiguously.
But the difference may explain the different resulting metallicity.
\citet{levshakov+2009} derive $\mathrm{[C/H]} = 0.1\dots 0.2$ and $\mathrm{[Si/C]} = -0.3$, indicating a moderate metal-rich absorber. 
While \citet{levshakov+2009} also apply a hard but slightly different radiation field, our very high metallicity of $\mathrm{[C/H]} \sim +0.8$ may be due to the underestimated \ion{H}{i} column density.

Finding silicon depleted in intergalactic absorption systems like for the systems at $z = 1.7529$ and $z = 2.8266$ is rather unusual.
According to \citet{aguirre+2008}, [Si/C] is expected to be $\sim +0.7$ indicating enrichment by Supernovae type II explosions.
However, studying metal-rich systems \citet{levshakov+2009} recently find that these absorbers often have underabundant silicon with respect to carbon.
The authors conclude that silicon is depleted into dust and, therefore, the absorber may by enriched by gas expelled from massive AGB-stars.

\section{Summary and conclusions}\label{conclusions}
 
We present a new, straightforward method to constrain the spectrum of the intergalactic UV background by means of photoionization modeling of metal line systems.
This new method requires simple, one-component systems with unblended, unsaturated absorption features of various species. 
In particular, at least two different ionization stages of two different elements are needed.
Then the column density ratio of two species of one element is used to construct a photoionization model of the system, while the second column density ratio is adopted to estimate the quality of the models for different ionizing spectra.
Based on the UV background according to \citet[][HM01]{haardt+2001} we parameterize three main characteristics of the spectrum:
the \ion{He}{ii} re-emission peak at 3\,Ryd, $\Delta_1$, the intensity level in the range 3-4\,Ryd, $\Delta_2$, which is related to the absorption due to the \ion{He}{ii} Lyman series, and the break at 4\,Ryd, the ionization edge of \ion{He}{ii}, $\Delta_3$.
The three parameters are measured in terms of deviation from the HM01 spectrum in dex.

The method is tested with artificially constructed systems using \ion{Si}{iii}/\ion{Si}{iv} to establish the photoionization models and \ion{C}{iii}/\ion{C}{iv} to evaluate the various SEDs.
The main conclusions are:
\begin{itemize}
  \item The SED of the ionizing radiation cannot be reconstructed unambiguously.
    Instead, we find various spectra that reproduce the observed column densities to a $1\,\sigma$ confidence level.

  \item The characteristics of the recovered SEDs are correlated.
    Evaluating the $1\,\sigma$ confident models yields that $\Delta_1$ is correlated with $\Delta_2$, i.e.\  the more pronounced the 3\,Ryd Ly$\alpha$ peak the more depressed is the intensity between 3-4\,Ryd.
    Therefore, we introduce $(\Delta_1 - \Delta_2)$, which measures the break at 3\,Ryd relative to the HM01 spectrum.
    Furthermore, $(\Delta_1 - \Delta_2)$ is correlated with $\Delta_3$, which determines the hardness of the spectrum.
    This means, harder SEDs favor a more pronounced break at 3\,Ryd.

  \item Nevertheless, the main features of the ionizing spectrum can be recovered by estimating the average value of $(\Delta_1 - \Delta_2)$ and inserting it into the $(\Delta_1 - \Delta_2)$-$\Delta_3$-relation.
\end{itemize}

Moreover, we investigate the dependence of the estimated physical parameters of the metal line system, i.e.\ the ionization parameter $\log U$, the metallicity [Si/H], and the relative abundance [Si/C], on the spectrum of the ionizing radiation.
\begin{itemize}
  \item The ionization parameter $\log U$ strongly depends on the 3\,Ryd Ly$\alpha$ peak $\Delta_1$, where higher $\Delta_1$ leads to lower $\log U$.
    This is due to adopting the \ion{Si}{iii}/\ion{Si}{iv} ratio to constrain the photoionization model since the ionization potentials of \ion{Si}{iii} and \ion{Si}{iv} are close to 3\,Ryd ($2.46\,\mathrm{Ryd}$ and $3.32\,\mathrm{Ryd}$, respectively) making both column densities sensitive to the spectrum in this energy range.
  \item The metallicity correlates with the 4\,Ryd break $\Delta_3$, confirming the previous result that harder ionizing spectra lead to higher metallicities.
  \item Considering all SEDs that reproduce the observed column densities within $1\,\sigma$ confidence, the resulting values of the physical parameters are spread over $\Delta\log U \sim 0.5\,\mathrm{dex}$, $\Delta\mathrm{[Si/H]} \sim 1.5\,\mathrm{dex}$, and $\Delta\mathrm{[Si/C]} \sim 0.5\,\mathrm{dex}$, respectively.
    Note, that in particular the uncertainty of the metallicity can be higher than one order of magnitude even for a well-fitting model if the spectrum of the ionizing radiation is unconfirmed.

    However, the uncertainties decrease if the hardness of the SED can be constrained.
    For a fixed 4\,Ryd break $\Delta_3$ the spread of $\log U$ and [Si/C] decrease by a factor of 2.
    In this case the spread of the metallicity decreases even more to $\Delta\mathrm{[Si/H]} \lesssim 0.5$ due to the strong dependence on the hardness of the ionizing spectrum.
\end{itemize}

The method is applied to three appropriate metal line systems at redshifts $z \approx 1.75$, $2.38$, and $2.83$.
The recovered spectra differ from the HM01 background yielding the parameters $(\Delta_1, \Delta_2, \Delta_3) = (+0.4, -0.5, +0.4)$, $(+0.4, 0.0, -0.3)$, and $(+0.3, -0.6, -0.4)$, respectively. 
However, the recovered SEDs are subject to high uncertainties.

(1) All systems favor an enhanced peak at 3\,Ryd suggesting that the \ion{He}{ii} recombination re-emission is stronger than predicted by the HM01 model.

(2) At $z < 2$ the recovered SED is rather flat and hard.
This is consistent with previous findings at similar redshifts by e.g.\ \citet{agafonova+2007}.
However, the adopted column density ratios are \ion{Si}{ii}/\ion{Si}{iii}/\ion{Si}{iv} and \ion{Al}{ii}/\ion{Al}{iii}, while for the $z>2$ systems we use \ion{Si}{iii}/\ion{Si}{iv} and \ion{C}{iii}/\ion{C}{iv}.
Therefore, the hardening of the UV background at lower redshifts should be confirmed using UV data where the \ion{C}{iii} $\lambda 977$ line can be detected even at $z < 2$.

(3) The recovered SEDs for the $z > 2$ systems tend to be softer than the HM01 background yielding a stronger break at 4\,Ryd in combination with enhanced \ion{He}{ii} recombination at 3\,Ryd and intensity depression at 3-4\,Ryd at least at $z \approx 2.83$.
Ionization of the observed systems by a soft radiation field implies that the absorbers contain much \ion{He}{ii}. 
Indeed, the expected \ion{He}{ii}/\ion{H}{i} ratios are high, $\log\eta \sim 2.4$ and $2.8$ at $z \approx 2.38$ and $2.83$, respectively.
This may suggest, that metal line systems reside in dense environments, where they are shielded from the general UV background and are exposed to a filtered radiation field \citep{miralda2005, schaye2006}.
In particular, the system at $z = 2.3799$ towards HS~1700+6416 is located within a proto-cluster of galaxies reported by \citet{steidel+2005}.
Generally, metal line systems are thought to arise close to galaxies where the heavy elements are produced and then transported into the IGM by galactic winds.
If this is true for the majority of metal line absorbers in QSO spectra, the usually adopted HM01 UV background might be too hard.
Due to the strong dependence of the metallicity estimate on the hardness of the ionizing radiation, the derived metallicities might be overestimated by up to one order of magnitude.
However, the estimated metallicity of the diffuse IGM \citep[e.g.][]{schaye+2003, aguirre+2008} is supposed to not be affected substantially. 

(4) Comparing the best-fit models adopting our recovered SEDs to models derived with theoretical UV background spectra from the literature we find a surprisingly well fit in case of the sawtooth modulated UV background from \citet{madau+2009} assuming delayed helium reionization.
\citet{madau+2009} provide three different models boosting the \ion{He}{ii}/\ion{H}{i} ratio by a factor of 10, 30, and 500 with respect to the standard case.
The boosting factor of the best-fit model increases with increasing redshift.
From direct observations of the \ion{He}{ii} absorption it is known, that it shows a Ly$\alpha$ forest structure and helium is therefore highly ionized at $z \lesssim 2.7$ \citep{kriss+2001, zheng+2004, fechner+2006b, shull+2010}.
Thus, the fraction of \ion{He}{ii} at these redshifts should be lower.
However, the observed systems are probably exposed to a surprisingly soft radiation.
They may reside in dense environments where remaining patches of \ion{He}{ii} might filter the UV background locally.

The application of our new method to three observed systems shows that it is possible in principle to infer constraints on the spectrum of the UV background.
The recent finding of a rather soft ionizing radiation should be confirmed with a larger sample of systems in the future.
However, systems exhibiting pronounced metal line features are probably residing close to galaxies.
They may not be representative for the general intergalactic medium since they live in special environments and may be affected by the radiation of local sources.
Thus, they may be exposed to a softer radiation than the general UV background.
As a preliminary result from the limited sample of systems investigated so far, we suggest that the \citet{madau+2009} spectrum with sawtooth modulation and delayed helium reionization is appropriate to study metal line systems.
In comparison, the spectrum provided by \citet{faucher+2009} seems to be too flat and featureless although it might be a appropriate representation of the mean UV background in intergalactic space.

\begin{acknowledgements}
We thank Dieter Reimers and Philipp Richter for valuable comments on the manuscript and Francesco Haardt for providing the MH09 spectra.
\end{acknowledgements}

\bibliography{/home/pollux/cfech/text/bibtex/papers}

\end{document}